\documentclass{article}
\usepackage{authblk}
\usepackage[utf8]{inputenc}
\usepackage{color,soul}
\usepackage{graphicx}
\usepackage{appendix}
\usepackage{amsmath}
\usepackage{amssymb}
\usepackage{amsthm}
\usepackage{comment}
\usepackage{caption}
\usepackage[ruled,vlined,linesnumbered]{algorithm2e}
\usepackage[margin=3cm]{geometry}

\usepackage{setspace}
\usepackage[section]{placeins}

\makeatletter
\AtBeginDocument{%
  \expandafter\renewcommand\expandafter\subsection\expandafter{%
    \expandafter\@fb@secFB\subsection
  }%
}
\makeatother
\makeatletter
\AtBeginDocument{%
  \expandafter\renewcommand\expandafter\subsubsection\expandafter{%
    \expandafter\@fb@secFB\subsubsection
  }%
}
\makeatother

\graphicspath{ {tex/figures/} }

\title{Combining the Projective Consciousness Model and Virtual Humans for immersive psychological research: a proof-of-concept simulating a ToM assessment}

\author[1,2,3]{Rudrauf, D.\thanks{Corresponding author: david.rudrauf@unige.ch}}
\author[2]{Sergeant-Perthuis, G.}
\author[1,2]{Tisserand, Y.}
\author[1,2]{Monnor, T.}
\author[2,4]{Belli, O.}

\affil[1]{FPSE, Section of Psychology, University of Geneva, Geneva, Switzerland}
\affil[2]{Swiss Center for Affective Sciences, University of Geneva, Geneva, Switzerland}
\affil[3]{Computer Science University Center, University of Geneva, Geneva, Switzerland}
\affil[4]{Evolutio, Geneva, Switzerland}
\theoremstyle{definition}

\newtheorem*{defn*}{Definition}

\theoremstyle{plain}

\newtheorem*{prop*}{Proposition}

\newtheorem*{theo*}{Theorem}
 
\theoremstyle{remark}

\DeclareMathOperator{\argmin}{\operatorname{\mathbf{argmin }}}

\newcommand{\R}{\mathbb{R}}

\DeclareMathOperator{\DKL}{\operatorname{DKL}}
\DeclareMathOperator{\FE}{\operatorname{FE}}

\newcommand{\Path}{\mathcal{P}}

\begin{document}

\maketitle

\newpage

\tableofcontents

\newpage

\begin{abstract}

Relating explicit psychological mechanisms and observable behaviours is a central aim of psychological and behavioural science. One of the challenges is to understand and model the role of consciousness, and in particular its subjective perspective as an internal level of representation, including for social cognition, in the governance of behaviour. Toward this aim, we implemented the principles of the Projective Consciousness Model into artificial agents embodied as virtual humans. Our goal was to offer a proof-of-concept, as a basis for a future methodological framework, aimed at simulating behaviours and assessing underlying psychological parameters, in the context of experiments in virtual reality, which would integrate processes emphasized by consciousness research. As an illustration of the approach, we focused on simulating the role of Theory of Mind (ToM) in the choice of strategic behaviours of approach and avoidance to optimise the satisfaction of agents' preferences. We designed an experiment in a virtual environment that could be used with real humans, allowing us to classify behaviours as a function of order of ToM, up to the second order. We show that our agents demonstrate expected behaviours with consistent parameters of ToM in this experiment. We also show that the agents can be used to estimate correctly each other order of ToM. A similar approach could be used with real humans in virtual reality experiments not only to enable human participants to interact with parametric, virtual humans as stimuli, but also as a mean of inference to derive model-based psychological assessments of the participants.    

\end{abstract}

\section{Introduction}

Human psychology entails highly complex information processing which plays a causal role in the generation of behaviours. Modelling such complexity to simulate human experience and behaviours, and predict outcomes of experimental research, is important for the development of psychological and behavioural science. An outstanding issue, and theoretical, methodological and technical challenge, is to understand and model how consciousness, and in particular its subjective perspective, may contribute to this process. 

Our approach stems from the rationale that computational models of human psychology should strive to be:
\begin{enumerate}
\item Integrative, i.e. targeting a comprehensive model of the human mind, including simulations of mechanisms related to consciousness and its subjective perspective, and their relations to perceptual, affective and social cognitive processes, 
\item Generative of embodied states and behaviours that could be measured in human participants in well-controlled and effective experimental and observation contexts, such as Virtual Reality (VR);
\item Capable of making inferences to assess internal psychological parameters in others based on their behaviours and through interactions with them, as a function of model parameters;
\end{enumerate}

Model-based agents embodied as Virtual Humans (VH) could then be used simultaneously: 

\begin{enumerate}
\item to serve as artificial confederates interacting with other virtual and/or real humans in the context of social psychology experiments;
\item to explore and test hypotheses about the mechanisms underlying observable behaviours (the question addressed with the approach would be for instance, which parameters can make my model behave and perform in a task as human would?) 
\item to serve as a tool to generate model-based psychological profiles and assessments (the question addressed with the approach would be for instance, can my model predict and explain observed interindividual differences in behaviour and performance in a given task?)
\end{enumerate}

Here, we present, in a preliminary manner, the approach we are developing for this purpose based on the Projective Consciousness Model (PCM). Our goal in this report is to present a proof-of-concept of the approach, using a limited example targeting the assessment of Theory of Mind (ToM), in a simple entry game, based on simulations in 3-dimensional virtual environments (VE) that could be used to run experiments in VR. We used a mock-up experiment designed for the purpose of demonstrating: 1) that our model can generate behaviours that would be expected from humans in a ToM task that could run in VR, and 2) that it is able to estimate the ToM parameters driving the behaviours of another artificial agent in the same task, which could be replaced by a real human (RH) in the context of an actual experiment.

\section{Background and rationale}

\subsection{General considerations about immersive environments and virtual agent modeling for psychological research}\label{intro-immersive}

The use of immersive virtual environment technologies has been proposed as a promising tool for social psychological research, capable of mitigating issues with experimental control-mundane realism trade-off, lack of replication, and nonrepresentative samples \cite{blascovich2002immersive}. In this perspective, Virtual Confederates, e.g. real humans embodied as virtual avatars, have been used as a research tool for overcoming limitations of real interactions with human confederates and paper-and-pencil designs \cite{de2014using}. Likewise virtual humans in gamified environments have been applied, sometimes in combination with machine learning analysis of human participants’ responses, to the screening of PTSD and other psychiatric disorders, through verbal interviews and the analysis of both verbal and non-verbal cues \cite{wortwein2017really, devault2014simsensei}, as well as to practicing negotiations \cite{kim2009bilat}. 

While certain approaches depart from it \cite{devault2014simsensei}, some have proposed to leverage game-theoretic frameworks combined with computational modeling of agents \cite{camerer2003behavioural}, integrating social utility functions, reasoning about iterative thinking limits, and statistical approaches, considering situations in which a player choice affects the payoff of other players, entailing a complex process of mutual influence and inference. Among different challenges for the development of models, a central one is the integration of simulations of mental representations that capture internal processes as they operate in human psychology: “Theorists analyze games in the form of matrices or trees but players presumably construct internal representations that might barely resemble matrices or trees” \cite{camerer2003behavioural}. We hold that this issue also entails to understand and model consciousness. 

\subsection{Consciousness theories: the problem of the subjective perspective}\label{machine-consc}

Much cognitive processing is unconscious and consciousness is only the tip of the iceberg \cite{velmans1991human, kihlstrom1996perception,van2012unconscious}. Nevertheless, it remains a central component of human information processing, and it is thus important to integrate models of consciousness and its impact on decision-making in models that wish to mimic human processing. 

Theories and models of consciousness developed over the last three decades encompass five broad, non-mutually exclusive conceptual frameworks \cite{reggia}: integrated information theories \cite{tononi2016integrated}, global workspace theories \cite{baars, dehaene2017consciousness}, internal self-model theories \cite{seth2012interoceptive}, higher-level representations, and attention mechanisms; the two first frameworks being the most prominent.

Overall, it can be said that consciousness operates as a global workspace with limited capacity, which accesses multimodal information, integrates it with memory, integrates mechanisms of uncertainty monitoring and reduction, and error corrections, to perform planning, decision-making and action programming, in a serial manner, through non-social and social imaginary simulations and appraisal \cite{baars, dehaene2017consciousness}. Along these lines, five ``axioms" have been proposed for artificial models of consciousness by Aleksander \cite{aleksander}: 1) presence, including mechanisms for representing the situated individual within the world; 2) imagination, of internal simulations of action without sensory input; 3) attention, to guide perception and modulate imagination; 4) planning, through the imaginary exploration of possible actions; 5) emotion, to evaluate plans. Furthermore consciousness integrates a representation of the body in space in relation to its environment, playing a role in homeostasis, survival and well-being, and relying on embodied appraisal and emotion \cite{seth2012interoceptive,damasio1999feeling,blanke2012multisensory}.

It remains largely unaddressed however, in particular from a modeling standpoint, how information is accessed, shaped and exploited through the global workspace of consciousness to accomplish the functions ascribed to consciousness. The issue directly relates to another essential axiom about models of consciousness that could be added to Aleksander's list: its qualitative experience or subjective character \cite{chella,manzotti}. For long, consciousness research has emphasized the phenomenologically pervasive and central role of a "subjective perspective", conceived of as a non-trivial, viewpoint-dependent, unified, embodied, internal representation of the world in perspective \cite{james1890principles, nagel1974like, lehar2003world, merker1, tononi2016integrated, rudrauf4}, from which different perspectives can be taken through imagination or action, in order to evaluate affordances and maximize utility \cite{riva1, rudrauf4, mchugh1}. The subjective structure in question would entail the combination of cognitive (spatial) and affective representations, for action programming \cite{ciompi1991affects, baroncohen3}. In complex social animals, perspective taking is also pivotal to perform theory of mind (ToM) \cite{Premack1}. Understanding and operationalizing how such subjective perspective, and how the appearances of contents in a perspectival non-Euclidean representation of the world may participates in the process of information integration and behavioral control carried out by consciousness is an important challenges for consciousness modeling \cite{seth,seth2,seth3,revonsuo,chella,manzotti}. While acknowledging the importance of the issue many have decided to set it aside \cite{dehaene2017consciousness, crick1990towards}. Others have proposed to address the issue based purely on information theoretic concepts, but largely fail to capture the phenomenon explicitly and in a specific manner as a result \cite{tononi2016integrated, merker2021integrated}. 

\subsection{The projective consciousness model and active inference}

The Projective Consciousness Model (PCM) \cite{rudrauf4,williford,rudrauf5,rudrauf3} aims at tackling explicitly the problem of the subjective perspective of consciousness and its role in active inference. 

Active inference conceptualizes the operation of the mind as a recursive cycle, including two main steps: 1) the inference of the causes of sensory information, 2) the planning of action. Resulting action outcomes provide a new context for the next cycle \cite{friston3}. Active inference has been formulated within variational optimization approaches, such as the Free-Energy Principle (FEP), which approximates Bayesian inference based on the minimization of a free energy acting as a cost function \cite{apps,friston2,friston3,limanowski,seth4}. Free energy is an upper-bound on surprise as a deviation between prior expectations and sensory evidence. Importantly, prior beliefs in an agent performing active inference can include models of preferences and desires, encoded as expectations. The approach has shown promises to understand the emergence of affective, affiliative and communicative behaviours \cite{friston5,constant,constant2,veissiere2020thinking,joffily,rudrauf3}. 

The PCM is based on two main principles: 1) consciousness is central to active inference in humans, 2) consciousness integrates information in a viewpoint-dependent manner, within a Field of Consciousness (FoC) in perspective, which acts, both in perception and imagination, as a global workspace \cite{baars,dehaene2017consciousness} and is governed by 3-dimensional projective geometry \cite{rudrauf4}. We recently showed how the PCM could explain and predict perceptual illusions such as the Moon Illusion, based on the calibration of a 3-dimensional projective chart under free energy (FE) minimisation, combining simulations and VR (\cite{rudrauf5}, see also \cite{rudrauf4}). But the FoC is thought to play a much broader role beyond perceptual experience. It corresponds to a 3-dimensional projective space, representing, within a subjective frame in perspective, an internal world model. One of its function according to the theory is to assess the distribution of affective and epistemic values ascribed to entities and actions in that world, as a function of perspectives being taken. It orders entities according to a point of view, which modulates their apparent size and thus relative importance. It can take multiple perspectives on the world model using projective transformations. Free energy can be expressed as a function of the FoC and thus of perspective taking, and reflects perceived or expected deviations from preferred values, as well as uncertainty with respect to sensory evidence. The idea is that its minimization through recursive imaginary perspective taking should make agents search for perspectives on the world that maximize their preferences and minimize uncertainty, and provide the agents with possible paths of action. We recently applied these principles to simulate complex adaptive and maladaptive behaviours among artificial agents, in a robotic context \cite{rudrauf2020role}. The agents embedded multi-agent models to infer the preferences of other agents based on their emotion expression and orientation, and to simulate other agents' FoC, according to projective transformations that respected known psychophysical laws. The process enabled the agents to appraise and predict other agents' behaviours, and plan their action accordingly. 

\subsection{Theory of Mind}\label{tom-def}

One interesting issue is to undersand how the subjective perspective of consciousness could play a role in Theory of Mind (ToM). ToM is the ability to infer others' mental states, beliefs and desires, and predict their behaviours for instance for strategic planning, and relies on the integration and imaginary manipulation of cognitive and affective information \cite{kalbe1,baroncohen1,baroncohen3,wimmer1}.

ToM is often conceptualized within simulation theory, which entails that humans use their own cognitive and affective appartus to imagine themselves in the position of others and simulate their subjective experience and likely behaviors; a process that would underline empathy \cite{lamm, berthoz2010spatial}. ToM through perspective taking is considered as important for emotion regulation and social-affective development \cite{gross,clement, gergely,beckes,fonagy,kalisch, dawson1}. It entails a balance between reward-expectation and the cost of executive function \cite{Koechlin2007}. 
 
Different levels of ToM have been distinguished \cite{hadwin1}. $Level-1$ and $-2$ respectively correspond to what is also described in the literature as Visual perspective taking 1 and 2 (VPT1 and VTP2) \cite{hamilton1}: the ability to infer, respectively, whether an object can be seen from a given point of view, and whether the object would look different from different points of view. $Level-3$ concerns the understanding that knowing requires verification through direct sensory evidence (or uncertainty reduction). $Level-4$ and $level-5$ concern the ability to understand respectively true and false beliefs in others, to predict their behaviour. $Level-6$ concerns the ability to understand that others can themselves perform ToM. Levels $1-5$ correspond to so-called $first-order$ ToM, and $Level-6$ to  $second-order$ ToM. The notion of order of ToM can be generalized recursively to $third-order$ ToM, i.e. the ability to understand that others can perform ToM about the ToM of others, and so on to $n-order$ ToM, up to the maximal capacity of an individual.

ToM is often assessed through verbal tasks \cite{baroncohen2,leslie1}, which may be susceptible to different biases, and it is important to develop non-verbal tasks assessing ToM based on outcome behaviors \cite{onishi1}. 

\subsection{Theory of Mind in models of agents and game theory, perspectives for immersive approaches}\label{tom-models}

Concepts of ToM can be found in classical Belief-Desire-Intention (BDI) models of agents \cite{georgeff1998belief}, which themselves entail embedded appraisal models \cite{broekens2008formal}.

ToM has become of interest in game theory \cite{Camerer1997} to understand and model rationale and irrational strategic planning in human and non-human competitive contexts \cite{Alskaif2015, Freire2019}. It relates to strategic uncertainty in the face of social situations with dependence on others’ choices, and entails assessing subjective probabilities based on others’ behaviours, and the interaction between processes of decision-making under risk and higher order beliefs about others \cite{nagel2018neural}.

Models of rationale, multi-agent coordination in unpredictable environments, applied to 2-dimensional strategic games, have been introduced, which implemented inferences about ToM, and took into account decision under uncertainty about others’ beliefs, based on recursive nesting of models (or Recursive Modeling Methods), in order to maximize expected utility, with mechanisms of belief update maximizing predictive power about observed behaviours of other agents \cite{gmytrasiewicz2000rational, gmytrasiewicz1998bayesian}. Likewise, simple Bayesian models of inferences of agents' intentions have been introduced to predict navigation in mazes, based on probabilistic inverse planning for action understanding \cite{baker2009action}. Recursive modeling of multi-agent interactions under uncertainty, using internal models of other agents integrating models of their preferences, and a variety of social influence factors and biases (such as consistency, self-interest, speaker’s self-interest, trust, likability, and affinity) have been investigated, with the overarching aim of exploring possible outcomes of intervention strategies through simulations, e.g. in the context of bullying \cite{pynadath2005psychsim}.

Of note, Bayesian models may be limited by the need to formulate an exact probabilistic framework, modeling a specific class of tasks and contexts. Although curse of dimensionality makes it difficult to compute posterior distributions in such frameworks, variational methods exist to mitigate this problem, which makes them more tractable.

Models based on adaptive control theory and probabilistic learning of agents have been investigated, using classical game theory paradigms and game-theoretic metrics, as an alternative to approaches such as Deep Learning \cite{Freire2019}. The latter approaches may yield good predictive power, but are based on distributed models that may be difficult to interpret, as they operate in practice as a black-box, while models with parameters that can be interpreted along psychological dimensions are warranted for many applications and psychological science. For instance, Yoshida et al. \cite{Yoshida2008} proposed an approach, confronting simulations and empirical data, to assess orders of theory of mind implied by behaviors, in a 2-dimensional competitive digital board game between two agents. The simulations were based on a model inspired from Reinforcement Learning (RL), recursively evaluating strategic predictions as a function of orders of ToM by maximizing expected future reward. The approach was then used to assess ToM capactities in Autism Spectrum Disorder participants \cite{yoshida2010cooperation}.

One general issue is to devise innovative methods capable of adapting and performing in a variety of situations and contexts, which are not always specifically designed for a given game theoretical paradigm and associated metrics. Such methods should be able to simulate, predict and assess behaviors, such as approach and avoidance, joint attention, emotion expressions, or more generally navigation in a 3-dimensional world, as a function of cognitive and affective processes such as ToM, in a more ecological manner, i.e. in a manner that is closer to real-world human behaviors as observed in the field. A promising approach is to combine computational models of agents and virtual reality, for instance to assess ToM capacities through simulations and embodied interactions.

As hinted in section \ref{intro-immersive} above, Virtual reality (VR) offers a promising framework to study social interactions through immersive technologies \cite{bombari2015studying}, in virtual environments that may be shared by human participants and virtual humans (VH) \cite{bruneau2015going,perry2015virtual}. Narang et al \cite{narang2019inferring} developed an approach combining a Bayesian model of ToM applied to artificial agents and VR, which was used to infer the intention of action of human participants in VR, and to control the navigation and approach-avoidance behaviors of virtual humans (VH) in social crowds. The model was a simple model, making inferences based on observed proxemics and gaze-based cues. 

Importantly, none of the approaches reviewed above aimed at modeling internal levels and mechanisms of representation that would integrate a model of the subjective perspective of consciousness.

\section{Model}

\subsection{Presentation of the model}

The model we present, as a proof-of-concept, is an implementation of the PCM principles close to that we introduced in \cite{rudrauf2020role}. The approach has similarities with \cite{Yoshida2008}, but instead of considering the expectation of a reward function for multiple agents, we consider a mean of free energy quantities within each agent that takes into consideration the simulation of active inference in other agents. In other words, each agent embeds a multi-agent model or system \cite{nguyen2020deep} to simulate others. Agents must thus be able to reverse infer preferences and the order of theory of mind of other agents, while in \cite{Yoshida2008}, it was the policy of other agents that was inferred. More generally, our approach is quite close to Recursive Modeling Methods that have been proposed in similar contexts, including mechanisms of inferences about preferences of others and factors of social influences \cite{gmytrasiewicz2000rational, gmytrasiewicz1998bayesian, pynadath2005psychsim}. The innovation is that we integrate an explicit model of the 3-dimensional subjective perspective of consciousness in the process, performing functions ascribed to consciousness based on view-point dependent subjective parameters (see Section \ref{machine-consc} above), in a manner that entails affective and epistemic (curiosity) drives based on projective geometrical mechanisms, and applies it to control VH in virtual environments. 

\begin{figure}
    \centering
    \includegraphics[width=0.8\textwidth]{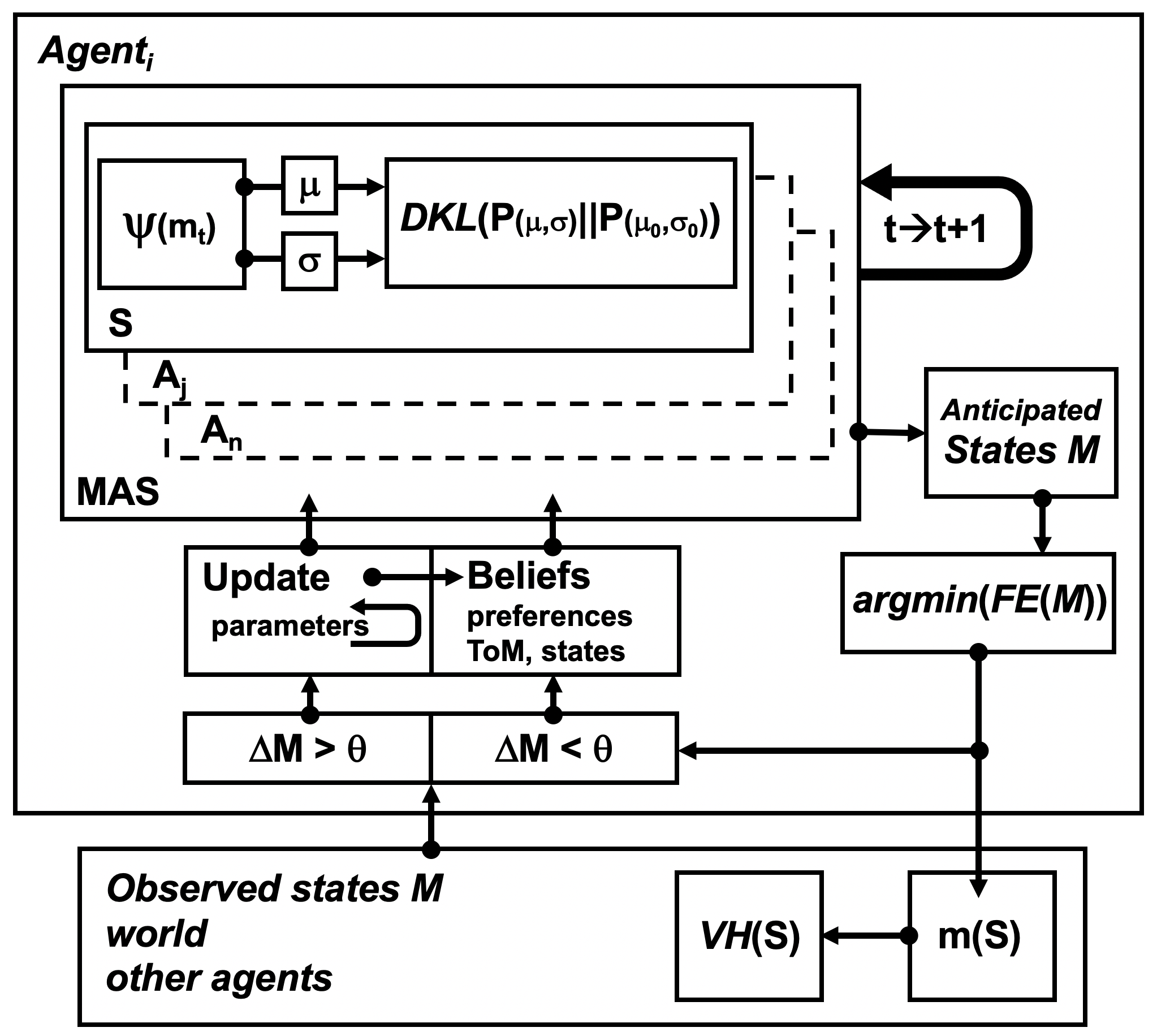}
    \caption{\textbf{Summary of the model architecture}}
    \medskip
    \small
    \raggedright
Each agent $A_i$ computes projections about itself as subject $S$, and about other agents $A_j$, using the same basic processing pipeline. For a given state or move $m_t$, evaluated by the agent, the agent computes a projective chart $\psi(m_t)$, corresponding to the FoC it attributes to a given agent, including itself. Perceived value $\mu$ and uncertainty with respect to sensory evidence $\sigma$ (given the current state of the agent) are computed based on $\psi(m_t)$ and the preferences attributed by $A_i$ to the agent under consideration. These parameters are used to define a parametric probability distribution $P(\mu,\sigma)$, which is compared to an ideal distribution $P(\mu_0,\sigma_0)$ through the Divergence of Kullback Leibler (DKL), yielding a cost function that is sensitive to divergence from both preferences and uncertainty. Emotions are also expressed by the agents accordingly (not indicated). The process is repeated recursively to assess successive moves, according to the depth of processing used by the agent (large round arrow, top right). The algorithm entails a Multi-Agent System (MAS) embedded within each agent. Multiple alternate sequences of moves $M$ are computed, to define a series of anticipated states. The sequence of moves that the agent retains corresponds to that which minimizes its overall free energy (FE), taking or not into account anticipations about other agents states (see text). The first move of the sequence is chosen by the agent as its actual move $m(S)$, which controls the state of the associated virtual human $VH(S)$. The agent then takes as inputs the observed states in the world, including of other agents (locations, orientations, emotion expressions). If those states diverge above a certain threshold $\theta$ from the anticipated states, a mechanism of reverse inference is triggered; otherwise, the agent keep computing projections based on its current beliefs, including preferences, ToM parameters, and more generally states (locations, orientations and emotion expressions of others). The mechanism of reverse inference tests different hypotheses about parameters such as preferences attributed to others and order of ToM used by others. It runs the same recursive algorithm used by the agent to simulate new projections, and retains the parameters that best explain the observed states in order to update its beliefs. 
    \label{fig:Fig_M1}
\end{figure}

We extend the previous version of the model with more advanced capacities of inferences, so that agents can infer preferences and ToM capacities in others, based on retrospective or prospective simulations of their behaviors, in a recursive manner. Predictions yielding best predictive power are used to update beliefs, in a manner that considers not only the emotion expression and orientation of other agents, but also their relative behaviors of approach and avoidance, as indicators of interest labeled with affective valence (see Figure \ref{fig:Fig_M1}).

When referring to active inference, we mean the process of inferring and acting according to inference recursively, which can be summarized as follows. Let $S$ be the space of sensory inputs and $\Gamma$ the space of states that the agent can be in, and let $M$ be the set of action the agent can perform. In the inference step a state $\gamma \in \Gamma$ is induced by sensory input $h$ by minimizing a cost function $c: S\times \Gamma \to \R$,

\begin{equation}
\gamma^{*}= \argmin_{\gamma\in \Gamma} c(h,\gamma)
\end{equation}

and during the action selection step, the subject chooses the action according to a second cost function $c_1: \Gamma \times M\to \R$, 

\begin{equation}
m^* =\argmin_{m\in M}c_1(m, \gamma^{*})
\end{equation}

which in turn induces a change at the level of the sensory input, since the environment reacts to this action.

In our setting we consider a collection of entities, $E$, constituted of objects and agents. Agents express emotions and can infer, and act according to their preferences and those ascribed to others, with respect to a situation, objects cannot act. When singling out an agent, for example when making explicit how active inference works for this agent, we will call it a subject. The space of agents will be denoted $A$. An agent $a\in A$ can express a positive emotion $e_{+}\in [0,1]$ and a negative emotion $e_{-}\in [0,1]$. The space of sensory inputs of a subject is constituted of the configurations of other entities in the ambient space and the emotions that agents express. The space of states is the preferences it can have for other entities when the subject does ToM-0, and preferences attributed to other agents for higher ToM. Subjects can act in two ways, they can move and express emotions.

The details of the following model we use are presented in \cite{rudrauf2020role}. 

The preference for an entity is a real number in $[0,1]$ denoted as $p$. Every subject, $s$, has an embodied perspective on the Euclidean ambient space that corresponds to a choice of a projective transformation that we denote as $\psi_s$; we will call it the projective chart associated to the agent. The quantity that links perspective taking and pleasantness of a situation is the perceived value $\mu$ that is computed for each entity $e\in E$ as,

\begin{equation}
\mu= p\gamma\frac{v_p^{1/4}}{v_{tot}^{1/4}}+ q_n(1-\gamma\frac{v_p^{1/4}}{v_{tot}^{1/4}})
\end{equation}

where $v_p$ is the perceived volume of the entity in the total Field of Consciousness (FoC) of the subject of volume $v_{tot}$. The perceived value $\mu$ is an average of the preference for the entity and a reference preference $q_n$ weighted by the relative perceived volume of the entity; the power $1/4$ on the volume is taken to match documented psychophysical laws. 

The subject also computes an uncertainty with respect to sensory evidence, denoted $\sigma$, that is greater with larger eccentricity with respect to the point of view of the subject, and the distance of the entity. In other words, there is more certainty about entities that appear actually or would be expected by imagination to be in front of and close to the subject. 

The subject is driven toward an ideal with high perceived value and low uncertainty. To compute the divergence from this ideal, the perceived value and uncertainty are associated to a probability distribution, $Q(.\vert \mu, \sigma)\in \mathbb{P}([0,1])$, centered in $\mu$ and of ``width" $\sigma$. This divergence is computed with the Kullback-Leibler diverge of $Q$ from the ideal distribution $P$ closely centered on values close to $1$. Let us recall that for any two probability distributions $P,Q\in \mathbb{P}(\Omega)$, over a space $\Omega$, with $dQ= f dP$,

\begin{equation}
\DKL(Q\Vert P)= \int f\ln f dP
\end{equation}

Let us now detail the active inference cycle of subjects with ToM of order $0$ (ToM-0) to ToM of order $2$ (ToM-2). Here, we shall not focus on the inference part of the process nor on emotion expression but rather on how agents select their moves, one can refer to \cite{rudrauf2020role} for a detailed presentation on how preference are updated and emotion expressed.

The preferences of a subject for the other entities with ToM-0 is encoded in a vector $(q_e,e\in E)$. The configuration of an entity, $e$, is a subset of $\R^3$ denoted as $X_e\subseteq \R^3$ and the collection of configuration will be denoted as $X$. The subject chooses its move $m$ form a set of moves $M$ by minimizing the following average of Kullback-Leibler divergences,

\begin{equation}
C_0(m,X,q)= \sum_{\substack{e\in E \\ e\neq s}} \frac{1}{\vert E\vert -1} \DKL(Q(.\vert \mu_{\psi(m),q}(X_e), \sigma_{\psi(m),q}(X_e)\Vert P) 
\end{equation}

When the subject performs T0M-1 (ToM of order $1$), it has a preference matrix $(p_{sae}\in [0,1],a\in A,e\in E)$ that encodes preferences that agents have with respect to other entities according to the subject.
The true preferences of the subject, i.e. the preference vector of $s$ is $p_{ss.}$. Agents may be influenced by other agents in the way they infer preferences and the way they act, this is respectively encoded by the influence vector on preferences $(J^p_{se},e\in E)$ and on moves $(J_{se}^m,e\in E)$. Subjects with ToM-1 can predict the move of the other agents assuming that they have order $0$ ToM; in fact they cannot assume that the other agents have a higher order of ToM or else it would contradict the fact that the subject has ToM-1. The number of steps in the future up to which the subject can predict the moves of the other agents is called the depth of processing and denoted as $dp$. At step $0$ of the prediction, the subjects attributes to an other agent, $a$, the preference vector $\tilde{q}^0_e=p_{sae}$ for entities $e\in E$; and the position of the entities is $X^0$. At step $k<n$ the predicted position, $X^k$, expressed emotions $e^k$ and preference vectors $\tilde{q}^k$ are used to predict the displacement, preference update and emotion expression of the others agents by applying active inference for ToM-0 as described in the previous paragraph; furthermore the subject has also an updated version of its preference matrix $p^k$. The subject then chooses its move at step $m^{k+1}$ by minimizing the following cost function, for $m\in M$,

\begin{equation}
C_1(m,p^{k}, J,Y^{k})=\underset{a\in A}{\sum}\underset{\substack{e\in E\\e\neq b}}{\sum}  \omega_{a,e}\DKL(Q(.|\mu_{a,\psi_a(m),p^{k}_{a.}}(Y^{k}_{e,m}), \sigma_{a,\psi_a(m)}(Y^{k}_{e,m})\Vert P)
\end{equation}

where $Y^{k}$ is the configuration of each entities that are not the subject at step $k+1$ and $Y^k_s$ is $X^k_s$. $Y^k_m$ is a mean to recall that the configuration of the entities depend on the move the subject decides to make, through $Y^k_{s,m}$. Here, for any agent $a\in A$ and entity $e\in E$,

\begin{equation}
\omega_{a,e}=J^m_{a}\frac{1}{\vert E \vert-1}
\end{equation}

One can remark that $C_1$ is in fact a weighted mean of several $C_0$. 

From this prediction $n$ steps in the future, the subject chooses the best set of moves that we assimilate to paths, $(m^{k*}_s, k\in [0,n])$, in a set of paths, $\Path$, by minimizing,

\begin{equation}\label{PCM:total-free-energy}\tag{FE}
\FE(m,p,J)=\underset{k\in [1,n]}{\sum}a_{k} C_s(m^k,p^k, J,X^k)
\end{equation}

where $\sum_{k=1...n} a_k=1$ and $a_k$ are chosen here to be $a_k=\frac{1}{n}$. The best move to make for the subject is the first move of the best path.

For a subject that has order $2$ of Tom (ToM-2), the same procedure as for a subject with ToM-1 holds: the subject can simulate the behaviours of the agent with respect to the degree of ToM it attributes to them, $(d_a,e\in A)$, and from these simulations it can decide what best sequence of moves to make. In order to do so, one should consider that the subject has a preference tensor $(h_{sabe}, a\in A, b\in A, e\in E)$ and influence matrices $(I^p_{sab},a\in A, b\in A),(I^m_{sab},a\in A, b\in A)$. The case we consider is simpler as we restrict the preference tensor $h$ to a preference matrix $p$, such as in ToM-1, by posing that $h_{sabe}=p_{sbe}$. When the subject starts its prediction of the behaviour of the other agents, i.e. at step $0$ of the prediction, the influence vectors of an agent $a$ believed to have ToM-1 by the subject are defined as $\tilde{J}^{0}_{a.}= I_{sa.}$. The cost function $C_3$ for the choice of the action of the subject at step $k$ of the prediction is a mean of the cost functions of the other agents depending on the degree of ToM that is attributed to them. We do not enter into more details on how $C_3$ is computed as in the experiment we consider, we assumed that the agents are not influenced in their action by how they believe other agents would feel as a result; what make the difference between a subject of ToM-1 and ToM-2 is how it predicts the behaviour of the agents, respectively attributing to them order $0$ or $0$ to $1$ of ToM.

\subsection{Inverse inference for ToM and preferences}

A subject with ToM-2 can attribute to another agent ToM-0 or 1 and for the subject to truly be able to perform ToM-2, it must be able to attribute correctly to the other agent the order of ToM it truly operates at. To do so, the subject must inverse infer the degree of ToM by analysing the behaviour of the agent.

When the prediction of the subject with respect to the actions of an agent diverge too much from its observed actions, it can start doubting its beliefs on the parameter it previously used in order to model the other agent's behaviour, and find better suited parameters; here the parameters are the preferences and the order of ToM attributed to the other agents (in the simulations below we will only use inverse inference on ToM order). To do so, the subject uses a measure of divergence from the predictions it made about the action of the other agent, $a^p$, with respect to the real action it has observed and memorized, $a$.

Let us consider the following example, at time $t$, the subject predicts the action $a^{p}(t)$ with respect to the information it holds about the preferences, $p(t)$, and order of ToM, $d(t)$, of the other agent. If the divergence, $f(a^p(t), a(t))$, is too large, then the subject will update the parameters of its model of the agent, i.e $p$ and $d$, in order to increase predictive power by minimizing a divergence,

\begin{equation}
(p^*,d^*)= \argmin_{p,d} f(a^p(p,d), a(t))
\end{equation}

The experiment we use below as a proof-of-concept is a two-choice simulation scenario. In this scenario, both the subject and another agent try to reach a vending machine among two, one being more attractive than the other, while trying to avoid running into each other. The subject cannot see the other agent before the near end of the experimental trial (except at the very beginning). The experiment is divided into two trials. A subject with ToM-2 knows that the other agent is also trying to avoid him and it is expected that if it encounters the other agent, it will learn from its mistake and revise the degree of ToM attributed to the other agent. Let us now explain how our agents revise their beliefs when confronted with evidence of misprediction.

A subject $s$ models the order of ToM of an agent by a random variable, that we denote as $D$, that takes two values $0$ and $1$. It has as prior law, $(p_1,p_0)\in \mathbb{P}(D\in \{0,1\})$ for $D$, that can be parameterized by $p_1$. The probability distribution plays the role of the belief the subject has on the order of ToM of the other agent. In the simulation scenario, the subject knows where the agent is at time $0$, but until the end of the trial, it does not have confirmation of its position; the subject speculates on the position of the agent at each time until it sees it, and gather new information on its position if it sees it eventually. To do so, it keeps in memory, the predicted position $(x_t^0,x_t^1)$ of the agent at each times $t$, respectively assuming that it performs ToM of order $0$ or $1$. At time $t+1$ if it does not see the agent then it predicts $x_{t+1}^0$ using the predicted position of the agent $x_{t}^0$ assuming that the agent acts as an agent with ToM of order $0$ and it predicts $x_{t+1}^1$ from $x_t^1$ assuming the agent acts as an agent with order $1$ ToM.

Therefore at each times $t$ the subject predicts two positions $x^0_t$ and $x_t^1$ for the agent. When the subject can attest the real position of the agent, it can confront it to the predicted positions. For example, if this assessment occurs at time $t_0$, the subject can confront its predictions with the true position of the agent, by considering $\vert x^0_{t_0}-x_{t_0}\vert$ and $\vert x^1_{t_0} -x_{t_0}\vert$, respectively the distance between the predicted position and true position of the other agent when the subject assumes that the agent has an order of ToM of $0$ versus $1$.\\

The subject computes

\begin{equation}
\mathbb{E}_{p_1}[\vert X_{t_0}-x_{t_0}\vert]= (1-p_1)\vert x^0_{t_0}-x_{t_0}\vert+p_1\vert x^1_{t_0} -x_{t_0}\vert
\end{equation}

as a metric for consistency of predictions; if this value is too high then the subject starts to doubt its priors and will look for $p_1$ that minimizes the previous quantity,

\begin{equation}
{p_1}_{t_0}^*=\argmin_{p_1} \mathbb{E}_{p_1}[\vert X_{t_0}-x_{t_0}\vert]
\end{equation}

One shows that this problem is the same as finding the minimum $d^*$,

\begin{equation}
d^*= \argmin_{d\in D} \left(\vert x^d_{t_0}-x_{t_0}\vert \right)
\end{equation}

as, ${p_1}_{t_0}^*=\delta(d^*)$, where $\delta(d^*)$ equals $1$ on $d^*$ and $0$ on the complementary.

\section{Virtual humans}\label{VH_section}

\begin{figure}
    \centering
    \includegraphics[width=0.6\textwidth]{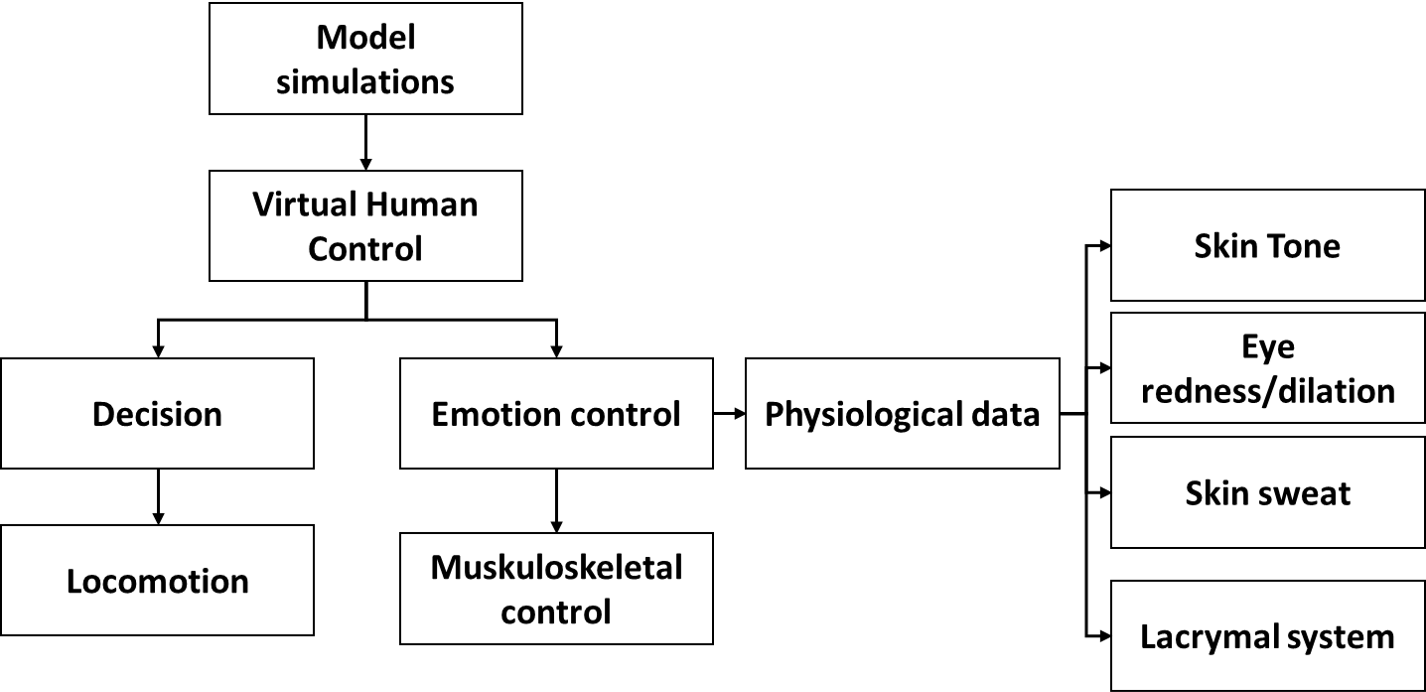}
    \caption{\textbf{Virtual Human control components}}
    \label{fig:Fig_VH}
\end{figure}

\begin{figure}
    \centering
    \includegraphics[width=0.8\textwidth]{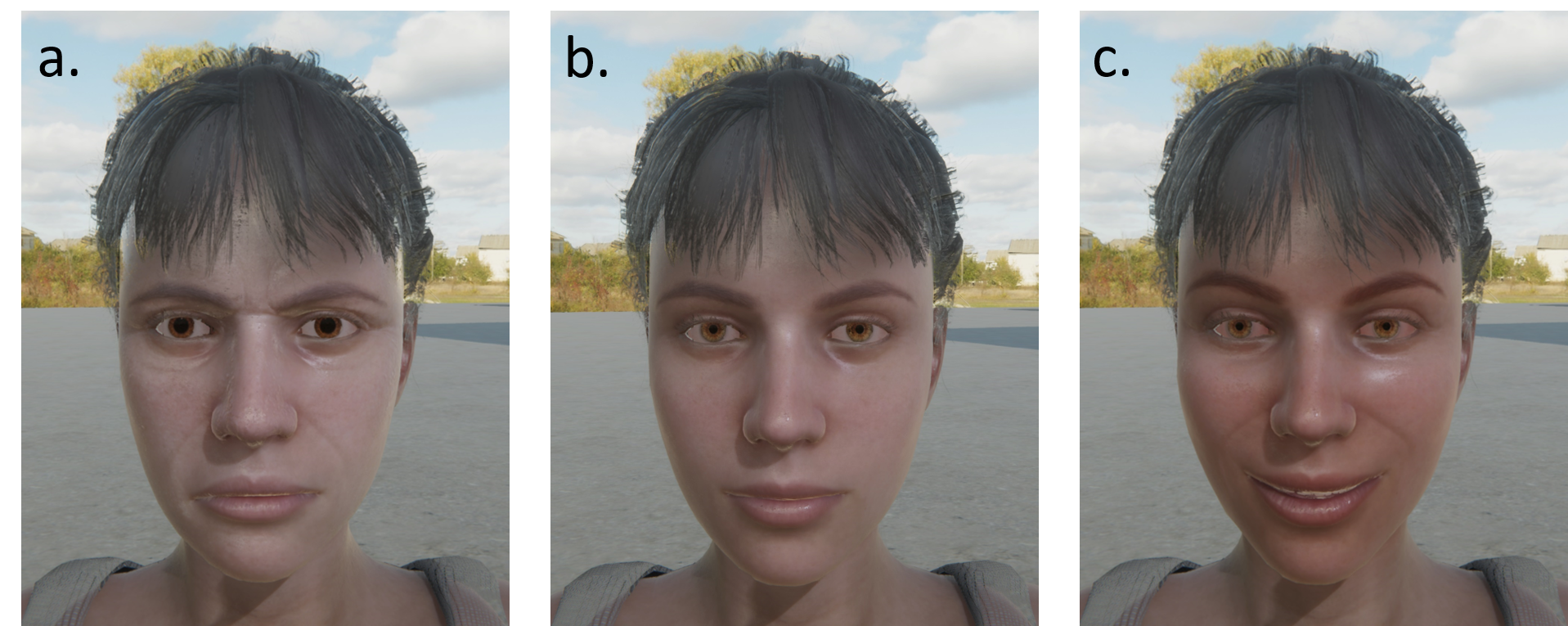}
   \caption{ Example of VH facial expressions including muskuloskeletal and physiological control}
    \medskip
    \small
    \raggedright
\textbf{a}. "High sympathetic" parasympathetic tone = 0, sympathetic tone = 100 (max), negative emotion expression; \textbf{b}. "Neutral": parasympathetic tone = 0, sympathetic tone = 0, neutral facial expression; \textbf{c}. "High parasympathetic", parasympathetic tone = 100 (max), sympathetic tone = 0, positive facial expression.
    \label{fig:Fig_face}
\end{figure}

Recently, we developed real-time simulations of VH with emotional facial expressions, combining physiological and musculoskeletal features \cite{tisserand2020real}. This VH control system is processing the data of our model simulations, to manage several facial expressions, including VHs musculoskeletal and physiological state (Figure \ref{fig:Fig_VH}).
The VH control system takes as inputs various information from simulations generated by the model, including the 3D world position of the agents, their successive positions in the virtual environment, their emotional states, the expected emotion during the next steps of the simulations, and their beliefs about other agents (positions, perceived values).
Based on this information, the VH system can procedurally generate a realistic visualization in which agents are moving and acting according to the simulations produced by the model.
A pathfinding system is used to apply smooth displacement to the VH during its navigation in the virtual environment based on the sequences of positions and orientations outputted by the model. This system is based on a state-of-the-art pathfinding method using Astar algorithm \cite{botea2004near}. The body movements are animated accordingly to be consistent with the overall movement of the VH. More precisely, the locomotion animations are blended based on the given direction and speed, which are passed as an input into a 2D cartesian mapping that controls animations’ transition.
Regarding facial expressions, the VH control system uses the emotion state of the agent, based on three main components: the positive and negative emotion intensity, and the surprise coefficient. Using these parameters, realistic facial expressions are generated and applied on the VH, including musculoskeletal control based on the Facial Action Coding System (FACS) \cite{ekman1997face}, and physiological control (skin tone, sweat, pupil dilation, eye redness) (Fig. \ref{fig:Fig_face}).
The musculoskeletal system is based on vertex displacement (blendshapes) and joint animation, while the physiological system is based on dedicated shaders developed for our VH system. The mapping between emotional states and VH parameters are summarized in table \ref{tab:VH-table}.

\begin{table}[]
\begin{tabular}{|lccc|}
\hline
                                                       & \textbf{Physiological} & \textbf{Musculoskeletal} & \textbf{Value}        \\ \hline
\textbf{Neutral}                                       & \multicolumn{1}{l}{}   & \multicolumn{1}{l}{}     & \multicolumn{1}{l|}{} \\ \hline
\multicolumn{1}{|l|}{Skin tone}                        & \multicolumn{1}{c|}{\checkmark} & \multicolumn{1}{c|}{}    & 0.5                   \\ \hline
\multicolumn{1}{|l|}{Pupil dilation}                   & \multicolumn{1}{c|}{\checkmark} & \multicolumn{1}{c|}{}    & 0.5                   \\ \hline
\multicolumn{1}{|l|}{Eye sclera and cornea redness}    & \multicolumn{1}{c|}{\checkmark} & \multicolumn{1}{c|}{}    & 0.5                   \\ \hline
\multicolumn{1}{|l|}{Skin sweat}                       & \multicolumn{1}{c|}{\checkmark} & \multicolumn{1}{c|}{}    & 0                     \\ \hline
\multicolumn{1}{|l|}{Facial expression}                & \multicolumn{1}{c|}{}  & \multicolumn{1}{c|}{\checkmark}   & 0                     \\ \hline
\textbf{Positive emotion with parasympathetic tone}    &                        &                          &                       \\ \hline
\multicolumn{1}{|l|}{Skin tone}                        & \multicolumn{1}{c|}{\checkmark} & \multicolumn{1}{c|}{}    & 1                     \\ \hline
\multicolumn{1}{|l|}{Pupil dilation}                   & \multicolumn{1}{c|}{\checkmark} & \multicolumn{1}{c|}{}    & 0                     \\ \hline
\multicolumn{1}{|l|}{Eye sclera and cornea redness}    & \multicolumn{1}{c|}{\checkmark} & \multicolumn{1}{c|}{}    & 1                     \\ \hline
\multicolumn{1}{|l|}{Skin sweat}                       & \multicolumn{1}{c|}{\checkmark} & \multicolumn{1}{c|}{}    & 0                     \\ \hline
\multicolumn{1}{|l|}{Facial expression (Action Units)} & \multicolumn{1}{c|}{}  & \multicolumn{1}{c|}{\checkmark}   & 6,12                  \\ \hline
\textbf{Negative emotion with sympathetic tone}        &                        &                          &                       \\ \hline
\multicolumn{1}{|l|}{Skin tone}                        & \multicolumn{1}{c|}{\checkmark} & \multicolumn{1}{c|}{}    & 0                     \\ \hline
\multicolumn{1}{|l|}{Pupil dilation}                   & \multicolumn{1}{c|}{\checkmark} & \multicolumn{1}{c|}{}    & 1                     \\ \hline
\multicolumn{1}{|l|}{Eye sclera and cornea redness}    & \multicolumn{1}{c|}{\checkmark} & \multicolumn{1}{c|}{}    & 0                     \\ \hline
\multicolumn{1}{|l|}{Skin sweat}                       & \multicolumn{1}{c|}{\checkmark} & \multicolumn{1}{c|}{}    & 1                     \\ \hline
\multicolumn{1}{|l|}{Facial expression (Action Units)} & \multicolumn{1}{c|}{}  & \multicolumn{1}{c|}{\checkmark}   & 1,4,15                \\ \hline
\textbf{Surprise}                                      &                        &                          &                       \\ \hline
\multicolumn{1}{|l|}{Facial expression (Action Units)} & \multicolumn{1}{c|}{}  & \multicolumn{1}{c|}{\checkmark}   & 1,2,5,26              \\ \hline
\end{tabular}
\caption{Virtual Human control mapping}
\label{tab:VH-table}
\end{table}

The following figure shows examples of facial expressions, generated using our VH control system, including muskuloskeletal and physiological features.

\section{Simulation of experiment}

We designed and simulated a social experiment to prove our concept. The experimental design was chosen so that behavioral outcomes would imply specific orders of ToM in participants performing the task. 

\subsection{Requirements}

We considered the following design and simulation requirements. First, the task behavioural outcomes should be unambiguous, and allow an experimenter to determine the order of ToM used by a participant who would follow the instructions of the task.  Second, the agents simulating participants in the experiment should demonstrate clear adaptive behaviours of navigation and emotion expression in a manner that is consistent with the task expectations, reflecting their attempts to optimise task outcome as a function of the order of ToM used by the agents to predict each other's actions and act accordingly. Here we considered agents that could be capable of order 0 (no ToM), order 1 (ToM at the 1st order), and order 2 (ToM at 2nd order) at most. Second, one agent in the experiment, the subject $S$, could be used as a virtual "psychologist" agent, and should demonstrate its capability to estimate the hidden ToM parameters of another agent $O$, based on its predictions and task outcomes. $O$ could then be conceived of, in this proof-of-concept, as a potential participant in an interactive experiment, designed to assess ToM capacities in the participant, and which could be implemented in VR. 

We reasoned that if PCM-driven VH $S$ could assess another PCM-driven VH $O$ correctly according to the experiment expectations, then they would be able to assess RH in an experiment with real human participants, either by interacting with them or by observing them. Our motivation was to render the simulation sufficient to prove the concept without actually running the experiment in real human participants, which was beyond the scope of this report.

\subsection{Experiment rationale and design}

\begin{figure}
    \centering
    \includegraphics[width=1\textwidth]{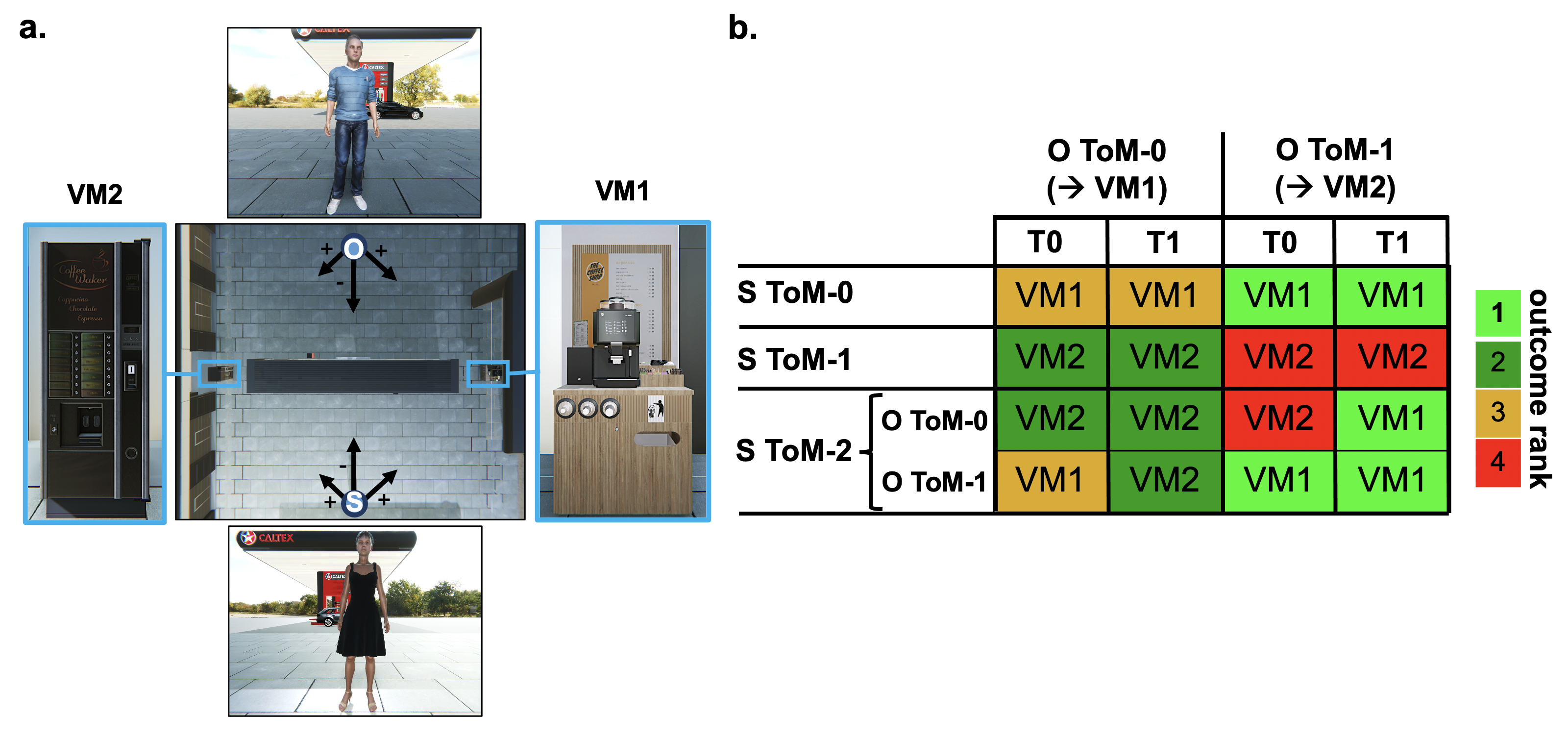}
    \caption{\textbf{Experimental setup and design}}
    \medskip
    \small
    \raggedright

\textbf{a.} Illustration of experimental setup. Two agents, a subject S and another agent O are on opposite sides of a small building (middle rectangle). On each side of the building, there is a vending machine, VM1 (right) and VM2 (left). Arrows arising from circles marking the initial position of the agents O and S, indicate fixed prior preferences towards entities: both S and O like (positive signs) VM1 and VM2 but prefer VM1 (longer arrow), and dislike each other (negative signs). \textbf{b.} The table shows the expected behavioural outcomes for the different combinations of orders of ToM at T0 and T1 for the subject $S$. $O$ can either perform ToM-0 (no ToM) or ToM-1 (first order ToM). If $O$ performs ToM-0, it should necessarily go to the preferred VW1, and if it performs ToM-1, to the less preferred VM2 (in order to avoid $S$ that it should expects at VM1). There are three possibilities of capacity for ToM for agent $S$: ToM-0, ToM-1, or ToM-2 (second order ToM). If $S$ is able to perform ToM-2, it can also perform ToM-1, and adjust the order of ToM it uses to optimize outcomes. Outcomes at T0 will depend on whether $S$ assumes initially that $O$ performs ToM-0 or ToM-1. At T1 agent $S$ should be able to revise its priors about the order of ToM of $O$, and optimise outcome on this basis. The experimental design is such that each row in the table is different from the other so that, when considering the different conditions of actual ToM order for $O$ (ToM-0 and ToM-1), and the two trials T0 and T1, the order of ToM that $S$ is capable of, as well as the order of ToM that $S$ attributes to $O$, can be determined. The table is color coded to indicate the rank of the outcomes in terms of optimisation of perceived value as a function of preferences (scale on the right), from best (rank $1$) to worst (rank $4$)(see text). 
    \label{fig:Fig_S01}
\end{figure}

We chose to design a rather simple and well-controlled entry-game, inspired by the common situations of having to choose between two stores or more generally destinations with different cost-benefits, in which one destination is more attractive than the other, but also more likely to be crowded with other people, which can hinder the free exploitation of the destination. The experiment was aimed to create a conflict of approach and avoidance based on non-social (intrinsic attractiveness of a destination) and social (social distancing and competition) preferences. Two agents, the subject S and another agent O, had to compete for reaching one of two vending machines (VM) selling coffee, on opposite sides of a building on a parking lot in a gas station. One of them, VM1 was better with higher quality products, and thus was the most attractive one for both agents. The other one, VM2, more mainstream, was not stocked with products of high quality, and thus was less attractive for both agents. Another assumption was that agents would prefer avoiding running into each other, which was operationalized as agents having negative preferences about each other. Both agents assumed that VM1 was most attractive to the other agent and VM2 a secondary choice (see Figure \ref{fig:Fig_S01}a). The situation was designed so that agents would have limited access to sensory evidence about their actual behaviour. Each agent could tell that the other was on the other side of the building at the beginning through sliding doors, first opened and then closed, and also assumed that the other agent was looking for a vending machine. However, a given agent would not be able to observe the behaviour of the other after the doors closed, except when reaching the areas of the vending machines on the side of the building. If both agents chose to go to the same vending machine, they could observe the other directly, if they chose to go to different vending machines, they could infer that the other was on the other side. 

The task for the agents was to choose which vending machine to go to (approach versus avoid), with the aim of maximising the satisfaction of their preferences. They could only use ToM (from order 0 to 2) to plan their actions. Only when reaching their destination could they use sensory evidence to attempt to infer the actual order of ToM used by the other agent, depending on the outcome and their capacity for ToM. The distance of the two vending machines from the agents at initial condition was equivalent in order to avoid a bias of distance on appraisal and the motivation of action, which would be entailed by the model, as in real situations. 

The “experiment” always involved the same initial setup with fixed preferences. Section \ref{simu_param} summarizes the fixed parameters of the simulation.

The possible outcomes of the experiment corresponded to decreasing levels of optimality in terms of satisfaction of preferences, e.g. level of free energy or emotion expressed, for the agent under consideration, from best to worst optimality rank, according to which: $rank 1$, $S$ finds itself alone at VM1 (thus $O$ went to VM2); $rank 2$, $S$ finds itself alone at VM2 (thus $O$ went to VM1; $rank 3$, $S$ finds itself in the presence of $O$ at VM1; and $rank 4$, $S$ finds itself in the presence of $O$ at VM2. 

Figure \ref{fig:Fig_S01}b presents the combination of experimental conditions and outcomes. The experiment combined three factors. The first factor, the Subject $S$ ToM capacity, corresponded to the maximum order of ToM for $S$, with three levels: ToM-0, ToM-1, ToM-2. When an agent was capable of ToM-2, it could also perform ToM-1 and potentially revise its beliefs about the order of ToM attributed to the other agent, based on outcomes. The second factor, the Other $O$ ToM capacity, corresponded to the actual order of ToM used by $O$, with two levels: ToM-0, ToM-1. This number of possible levels for factor two was prescribed by the fact that the experiment assumed the subjects' maximum order of ToM to be ToM-2. This is a logical consequence of the very definition of ToM: an agent with a maximum order of ToM $n$ (ToM-$n$) can at most make inferences about another agent with maximum order of ToM of $(n-1)$ (ToM-$(n-1)$) (see \cite{Yoshida2008}). Since the experiment was designed to assess a subject's maximum order of ToM up to ToM-2, the experiment had two possible actual orders of ToM for agent $O$: ToM-0 and ToM-1. This could be generalized to higher orders of ToM, which our model is capable of performing, but we wanted to limit the number of conditions of the experiment for the sake of the clarity of presentation. 
The third factor was the trial number, as the task was repeated twice for each condition, with two levels: T0, the initial action, and T1, the second attempt, so that subjects with sufficiently high capacity for ToM (ToM-2) could infer the ToM order of the other agent $O$, and adapt their behaviour for the second trial, to reach a more optimal outcome. 

This design guaranteed that we could estimate the order of ToM of an agent following the task instructions by comparing outcomes as a function of condition. It is important to note that during both trials, the order of ToM of $O$ is assumed to be fixed. 

The expected outcomes of the experiment are derived from the behaviour expected from a RH and are presented in Figure \ref{fig:Fig_S01}b. Let us explain how these outcomes are derived when $S$ can have a ToM of order $0,1,2$, and O of order $0,1$. If an agent performs ToM-0 then it goes toward its preferred machine $VM1$, if an agent performs ToM of order $1$ then it is expected to go to the VM opposite to the one the predicted agent with order 0 would go to.The first table correspond to the case where O performs ToM-0 and therefore always goes to VM1 and the second one to the case where it performs ToM-1 and goes to VM2. The experiment is divided into two times, trial $T0$ and $T1$, if S does not perform ToM-2 then it can't learn from its mistakes and the outcomes of both trials is the same; it is reported in the first two lines of the tables.
If $S$ performs ToM-2 it may initially assume that $O$ has ToM-0 (the less greedy hypothesis on $O$). Then we expect that after one trial of the experiment, $S$ would confront its expectations to the outcome of the trial. If it predicted correctly the outcome and does not meet $O$, then it does not have to change its beliefs and therefore may assume that $O$ had ToM-0 and is located around $VM1$, for both trials T0 and T1. Thus $S$ should go to VM2 in both trials. However, if the beliefs of $S$ were false during the first trial, both agents should arrive at $VM2$, and $S$ is expected to change its belief about the ToM order of $O$, from ToM-0 to ToM-1. In the next trial $S$ should go to $VM1$ and $O$ to $VM2$. A second possible case is if $S$ initially assumes that $O$ has ToM-1, in which case, similarly, $S$ would have to update its beliefs to account for the false prediction. The expected results of the experiment for a subject with ToM-2 are reported in the last two lines of the tables, the first one corresponding to the situation where the subject believes at first that O has ToM-0 and the second one ToM-2.

\subsection{VHs simulation procedure}

At the beginning of each simulated case, the positions of the agents, the divider between the two VHs and the coffee machine were procedurally set based on the given simulation. The visualization was created using Unity (v2021.1), with the High Definition Rendering Pipeline (HDRP). The VHs model was designed using Character Creator. The VH control was handled by the Geneva Virtual Humans toolkit \cite{tisserand2020real}, managing the locomotion, facial musculoskeletal and physiological parameters, as described in Section \ref{VH_section}. A preview window allows the real time visualization of simulations, and the offline rendering of the simulation with a multi-view system. The system allows the creation of 4K videos from various point of views, a radar view that shows the path of the agents, a general viewpoint of the simulation, a camera focusing on the facial expressio-ns of each VH, and the first-person views for each VH.
The scene is designed to run both in real-time desktop and VR applications and offline with up to 4K rendering. The scene is VR ready, for future real-time experiments in which a RH participant could be part of the scenario. The scene can already be observed by an experimenter within VR.

\begin{figure}
    \centering
    \includegraphics[width=0.8\textwidth]{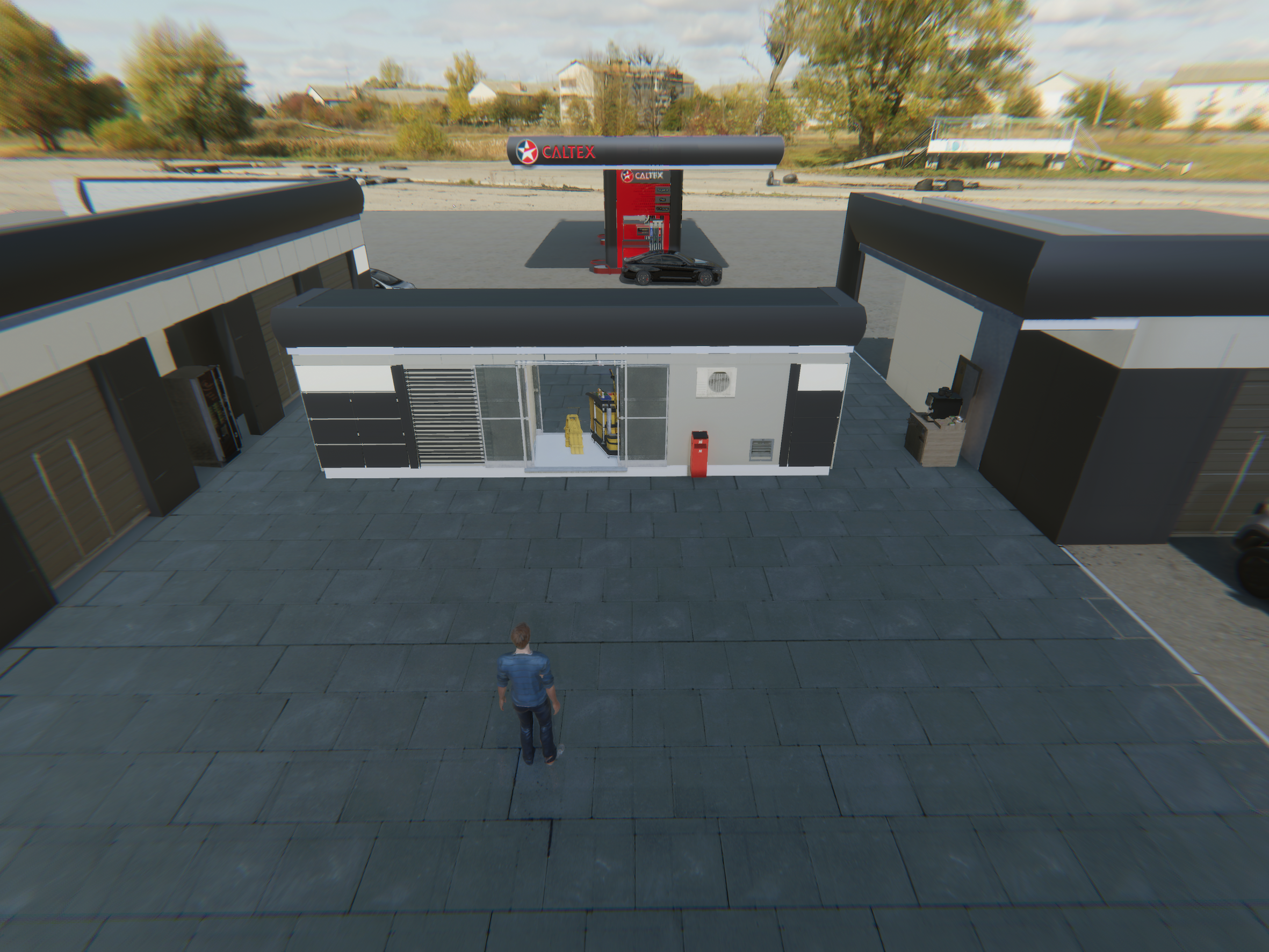}
    \caption{\textbf{Overview of the scene}}
    \medskip
    \small
    \raggedright
Screenshot of the simulation viewer.
    \label{fig:Fig_Viewer}
\end{figure}

\subsection{Simulation parameters}
\label{simu_param}

The subject is labelled $1$, the agent $2$, VM1 is $3$ and VM2 is $4$. In the experiment, the preference matrix for the subject, $p_1$, is the following,

\begin{center}
\begin{equation}
\begin{pmatrix}
0.5 & 0.01 & 0.6 & 0.55\\
0.01 & 0.5 & 0.6& 0.55
\end{pmatrix}
\end{equation}
\end{center}

The preference matrix of the other agent is,

\begin{center}
\begin{equation}
\begin{pmatrix}
0.5 & 0.01 & 0.6 & 0.55\\
0.01 & 0.5 & 0.6& 0.55
\end{pmatrix}
\end{equation}
\end{center}

These matrices contain more information than necessary when the subject or agent performs ToM-0 and the preference vectors in these cases are the true preferences extracted from the previous matrices.

The influence matrix of the subject for preferences and action (see above) are reduced to self influence, only $I_{saa}\neq 0$ and similarly for the influence vector of the other agent.   

\subsection{Results}

\begin{figure}
    \centering
    \includegraphics[width=1\textwidth]{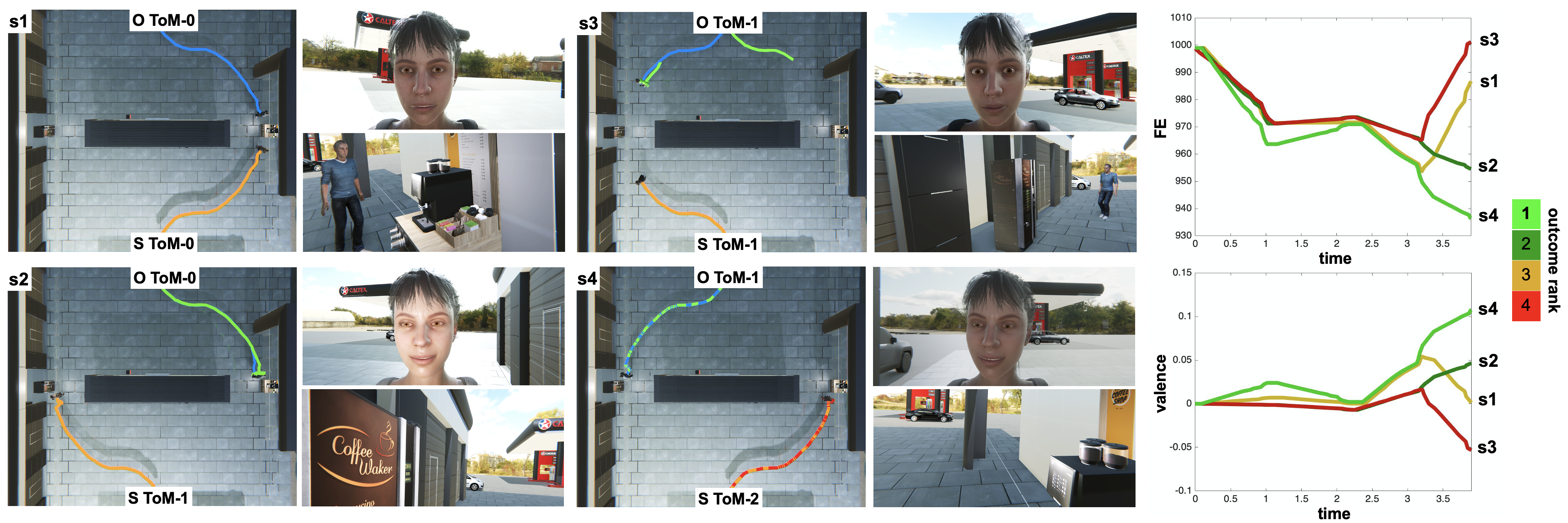}
    \caption{\textbf{Basic behaviours of the agents as a function of ToM}}
    \medskip
    \small
    \raggedright
     For each situation, s1, s2, s3, s4, images of virtual environment. \textit{Left}: view from above. Orange traces are $S$, blue traces are $O$, green traces are predictions about $O$ according to $S$, red traces are predictions about $S$ attributed to $O$ by $S$ (if $S$ uses ToM-2). \textit{Upper-right}: woman VH face close-up. \textit{Lower-right}:  first person perspective of $S$ on $O$ (male VH). \textbf{s1}. $S$ and $O$ perform ToM-0. Agents run into each other at VM1 . $S$ emotion expression is rather unhappy. \textbf{s2}. $S$ performs ToM-1 and $O$ ToM-0. Agents do not run into each other. $S$ goes to VM2. $S$ emotion expression is rather happy. \textbf{s3}. Both $S$ and $O$ perform ToM-1. Agents run into each other at VM2. $S$ emotion expression is unhappy. \textbf{s4}. $S$ performs ToM-2 and $O$ ToM-1. Agents do not run into each other. $S$ goes to VM1. $S$ emotion expression is happy. See free energy (FE), valence, and outcome rank in the charts on the right, corresponding to each situation as labeled.
    \label{fig:Fig_R1}
\end{figure}

We focus on the results regarding $S$. The results are consistent with the design. Agents behave as expected as a function of their relative ToM order. Their end state free energy (FE) and expressed valence are consistent with the corresponding outcome ranks. Figure \ref{fig:Fig_R1} shows results of basic behaviours of the agents as a function of ToM contingencies, without $S$ performing inverse inference to learn the order of ToM of $O$.

\begin{figure}
    \centering
    \includegraphics[width=1\textwidth]{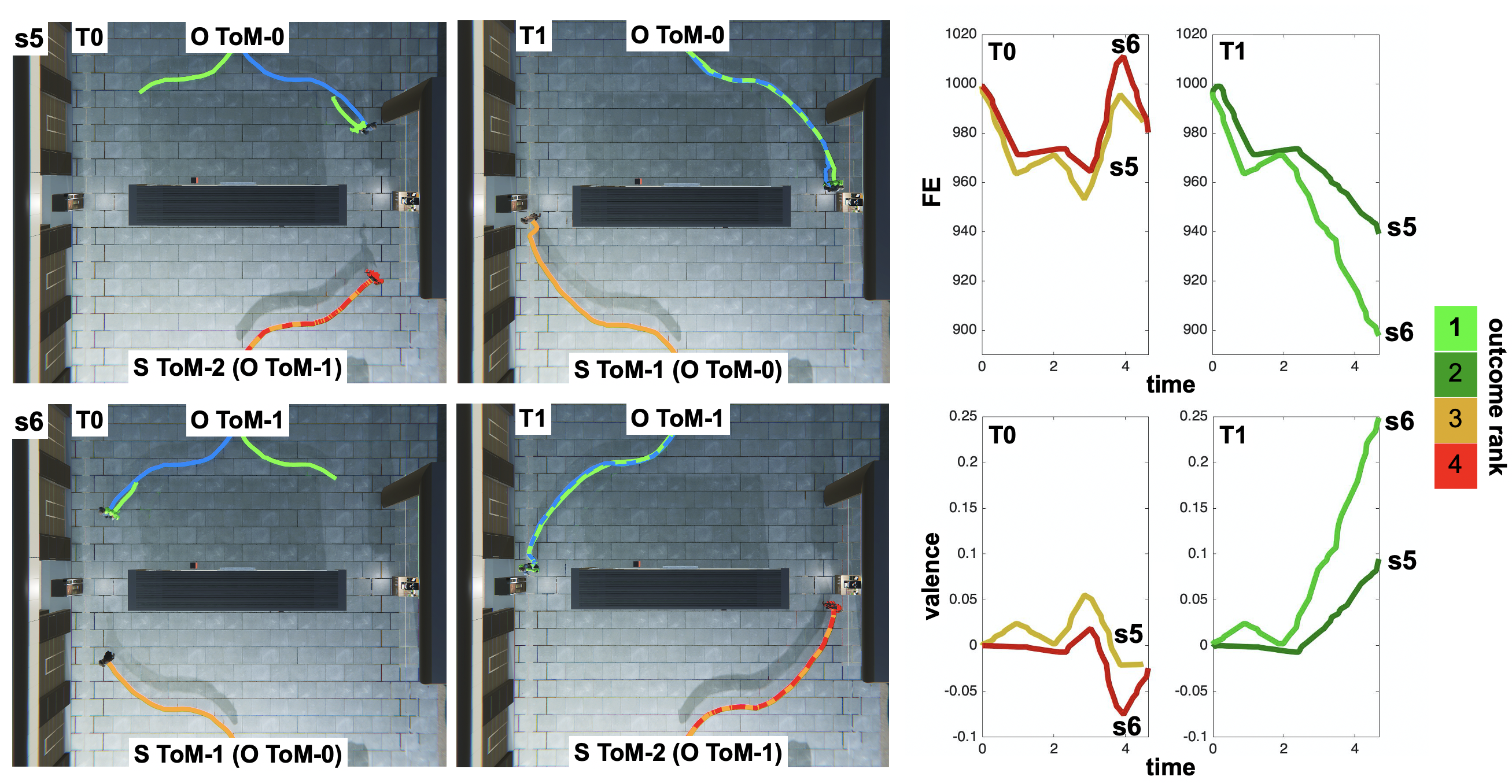}
    \caption{\textbf{$S$ performs inverse inference about $O$ ToM order}}
    \medskip
    \small
    \raggedright
    For each situation, s5 and s6, images of virtual environment at trial T0 and T1. Orange traces are $S$, blue traces are $O$, green traces are predictions about $O$ according to $S$, red traces are predictions about $S$ attributed to $O$ by $S$ (if $S$ uses ToM-2). \textit{Left}: view from above at T0. \textit{right}: view from above at T1. \textbf{s5}. T0: $S$ performs ToM-2 (and attributes ToM-1 to $O$) while $O$ actually performs ToM-0. Agents run into each other at VM1. T1: $S$ has inferred $O$ ToM order, and obtains a better outcome, finding itself alone at VM2. \textbf{s6\textit}. T0: $S$ performs ToM-1 (and attributes ToM-0 to $O$) while $O$ actually performs ToM-1. Agents run into each other at VM2. T1: $S$ has inferred $O$ ToM order, and obtains the best outcome, finding itself alone at VM1. See free energy (FE), valence, and outcome rank in the charts on the right, corresponding to each situation as labeled, for T0 and T1.
    \label{fig:Fig_R2}
\end{figure}

$S$ succeeds in inferring $O$ ToM order after the first trial (T0), and optimises outcome accordingly. Figure \ref{fig:Fig_R2} shows results of the behaviour of the agents as a function of ToM contingencies, when $S$ performs inverse inference to learn the order of ToM of $O$ at the second trial (T1).

\section{Discussion}

Our objectives in this report were to demonstrate: 1) that we could integrate an explicit model of the subjective perspective of consciousness in a recursive, multi-agent model (PCM-driven agents) that could be combined with virtual humans (VH), 2) to simulate behaviors of approach-avoidance, which would be expected to result from configurations of preferences and orders of ToM in real humans (RH), 3) in an experiment that could classify these behaviors in terms of underlying generative models as a function of orders of ToM, based on clear behavioral outcomes, 4) that could be implemented in virtual reality (VR) in the future for actual experiments with RH, and 5) that our PCM-driven agents could correctly infer parameters such as ToM order used by other agents, as a basis for the future use of such approach to assess RH in immersive experimental contexts such as VR. The approach leverages active inference and projective mechanisms to enable artificial agents to interact with each other in a 3-dimensional virtual environment. 

\subsection{Killing three birds with one stone}

There are three complementary points of view that can be taken on the experiment we presented as a proof-of-concept of an approach to the problem of integrating computational models, artificial agents and virtual humans, in a unified framework for psychological science. 

First, the experiment was a way of assessing whether model-based virtual agents, here using the PCM, would be capable of demonstrating behaviors that would be expected in the task, according to their order of ToM, as would be expected from a human performing the same task. Indeed we have built an experiment that was adapted \textit{a priori} for distinguishing RH behaviours, and we have shown that our VH are a good sampler of what would be RH behaviours with respect to this experiment, as they simulates accurately these expected behaviour. Even though we did not explicitly compare our results to empirical data in this contribution, generally speaking, our results are consistent with comparable approaches \cite{gmytrasiewicz2000rational, gmytrasiewicz1998bayesian, pynadath2005psychsim}, which however did not aim at modeling consciousness explicitly, and thus were not as much constrained by the aim of capturing internal processes as they operate in human psychology \cite{camerer2003behavioural}.

Second, we have also shown that the experiment is a good classifier for the behaviours of the VH as different parameter of the model that generate the behaviours of the VH can be distinguished by the experiment.

Third, the approach used agents that could make inferences about psychological parameters driving others, as a human experimenter might do to assess the determinants of observed behaviors in humans, i.e. as a way to test the ability of agents to correctly infer others' behaviours. The results demonstrated that the target parameters of the model, the order of ToM, could be well discriminated by the VH that we called subject $S$ in the experiment, since it was able to discriminate the order of ToM of the other agent (ToM-0 or ToM-1) consistently, based on its own internal mechanics of inverse inference as embedded within its recursive multi-agent architecture. Importantly, $S$ would be able to do so even if the agent was a real human agent in a VR experiment leveraging motion tracking information from the human participant. 

In other words, the same model that could be used to control artificial agents embodied as virtual humans in a VR task, could be used in the future to make inferences about the behaviours of real humans that would participate in the same task. The rationale is that such agents could then assist investigators with assessing and reporting hidden psychological parameters and profiles of different groups and individuals. Although very preliminary, this is in line with a research program with the overarching goal of building synthetic psychology, i.e. the rebuilding of human psychology into artificial systems. Indeed, if successful, such program should yields artificial agents that can interact with, and interpret each others as well as real humans, as we would expect from real humans toward each other, because they share similar internal representations and processing mechanisms. For simplifying the presentation and discussion, we limited the simulations to agents with at most ToM of order 2, but the framework can be used to simulate agents with higher-order ToM capacities. Thus one could easily extend the design of the experiment and add a third PCM-driven agent, which we could call the observer, and which would be capable of ToM of order 3, and thus be able to predict the degree of ToM of the agents up to the second order. Results in this direction are promising but beyond the scope of this report, and will be presented in a future report assessing TOM of order $n$ in groups of PCM-agents.

We used simulations of a simple entry-game, i.e. a very controlled, and somewhat narrow task, in order to demonstrate our proof-of-concept, i.e. by showing that the model could infer the order of ToM in others through interactions in a 3-dimensional virtual environment. We chose to focus on this particular illustrative example using simulations only, as a more systematic applications to a variety tasks, as well as systematic comparison with empirical data, was beyond the scope of this theoretical and technical, preliminary contribution. But the same model could be used to perform tasks in which agents would have to infer preferences in other agents based on their behaviours, which include their orientation or move towards or away from objects and others, and corresponding emotion expressions. Likewise, thanks to its operation, assuming adequate experimental and task design, the model could be used to infer a combination of parameters (preferences and order of ToM in others, or any other parameters that the model include). More generally, PCM-driven agents can be used to simulate behaviors in a variety of contexts (see for instance our recent work on simulating adaptive and maladaptive behaviors relevant to developmental and clinical psychology \cite{rudrauf2020role}). 

\subsection{Limitations and perspectives}

As this point, without tailored restrictions limiting computational load, the PCM algorithm does not run in real-time, and cannot be used in practice for many real-time interactive VR experiments. Optimisation approaches both at the software and hardware levels are currently being pursued in order to mitigate the problem. In order to close the loop of interactions between real humans and virtual humans in VR, parameters need to be extracted from VR participants to serve as inputs to the model, so that PCM-driven agents will interpret VR participants as they interpret other PCM-driven agents. This can be achieved by leveraging VR technology, including motion capture, eye-tracking and smart interfaces for emotion expression capture, but this important issue is beyond the scope of this contribution.

Another interest of the approach we develop is that it could be used to assess and optimize the capacity of different immersive experimental designs to discriminate psychological parameters and mechanisms based on behaviors (see also \cite{pynadath2005psychsim}). Different designs could be explored through mass simulations of PCM-driven agents along a variety of parameters, independently assessed or in combinations, e.g. preferences and orders of ToM. If target parameters can be well distinguished based on behavioral outcomes, the design would appear sufficient as a classifier. On the contrary, if different sets of parameters would predict the same outcome behaviors in an ambiguous manner, that would mean that the design of the experiment is not fully discriminant. Designs could then be revised and retested by adding conditions. For instance, if subsets of parameters could not be reliably inferred simultaneously by agents due to the ambiguity of behaviors, the experiment could introduce different phases of estimations in which the parameters could be assessed independently, e.g. by observing first the behaviors of agents without interacting with them to infer their mutual preferences, and then, based on this prior, attempting to infer their order of ToM through interactions with them. 

In this proof-of-concept we used fixed preferences that were identical between the two agents. This lead to perfect predictions of the trajectories of the other agent by a given agent, when the order of ToM was adequate. Using different values of preferences between those attributed to an agent and those actually used by the agent may have induced some error of prediction. Manipulating such differences and studying their impact on fine grained behaviours is interesting but beyond the scope of this contribution. Note however, that generally speaking the algorithm is robust, and slight randomizations of the preferences we performed did not lead to different behaviours that would have impacted the results of the experiments.    
Finally, let us note that the way a subject predicts the action of the other agent several steps in the future and uses this prediction in order to choose its best move can be seen as an optimal control problem with a cost function depending on time. It might be possible to consider that the whole simulated situation is itself a control problem, and optimal policies might give a more optimal update rule for simulating the action of the other agent by the subject. However, the approach we chose focused on modeling agents as subjective systems, who model and predict others' actions based on their own internal mechanisms of action selection, as prescribed by simulation theory \cite{lamm, berthoz2010spatial}, without having necessarily the possibility of developing a full representation of the problem for more globally optimal solutions.

\subsection{Conclusion}

The approach we developed combines PCM-agents, integrating a 3-dimensional model of the subjective perspective of consciousness, and virtual humans, with the overarching aim of studying human behaviours and their psychological determinants in virtual reality experiments. At this point, our implementation remains preliminary, calls for applications in more tasks and contexts than the narrow illustrative entry-game example we chose, as well as empirical validations. We believe however that the proof-of-concept we presented offers a promising ground toward that aim.  

\section*{Authors' contributions}
David Rudrauf: conceived and developed the principles of the model and algorithm, co-designed the proof-of-concept experiment, performed analyses of the simulation results, and co-wrote the article.  

Grégoire Sergeant-Perthuis: developed the mathematical formulation of the model and algorithm, co-developed aspects of the model, co-designed the proof-of-concept experiment, and co-wrote the article. 

Yvain Tisserand: helped with the conceptual and technical rationale, implemented the virtual humans and virtual environment, performed analyses of the simulation results, and co-wrote the article.

Olivier Belli: developed and implemented the code used for the simulations, ran the simulations, and co-wrote the article.

Teerawat Monnor: contributed to discussions about the design of the proof-of-concept, and to co-wrote the article.

\bibliographystyle{ieeetr}
\bibliography{bintro}

\begin{thebibliography}{10}

\bibitem{blascovich2002immersive}
J.~Blascovich, J.~Loomis, A.~C. Beall, K.~R. Swinth, C.~L. Hoyt, and J.~N.
  Bailenson, ``Immersive virtual environment technology as a methodological
  tool for social psychology,'' {\em Psychological inquiry}, vol.~13, no.~2,
  pp.~103--124, 2002.

\bibitem{de2014using}
C.~M. de~Melo, P.~J. Carnevale, and J.~Gratch, ``Using virtual confederates to
  research intergroup bias and conflict,'' in {\em Academy of Management
  Proceedings}, vol.~2014, p.~11226, Academy of Management Briarcliff Manor, NY
  10510, 2014.

\bibitem{wortwein2017really}
T.~W{\"o}rtwein and S.~Scherer, ``What really matters—an information gain
  analysis of questions and reactions in automated ptsd screenings,'' in {\em
  2017 Seventh International Conference on Affective Computing and Intelligent
  Interaction (ACII)}, pp.~15--20, IEEE, 2017.

\bibitem{devault2014simsensei}
D.~DeVault, R.~Artstein, G.~Benn, T.~Dey, E.~Fast, A.~Gainer, K.~Georgila,
  J.~Gratch, A.~Hartholt, M.~Lhommet, {\em et~al.}, ``Simsensei kiosk: A
  virtual human interviewer for healthcare decision support,'' in {\em
  Proceedings of the 2014 international conference on Autonomous agents and
  multi-agent systems}, pp.~1061--1068, 2014.

\bibitem{kim2009bilat}
J.~M. Kim, R.~W. Hill~Jr, P.~J. Durlach, H.~C. Lane, E.~Forbell, M.~Core,
  S.~Marsella, D.~Pynadath, and J.~Hart, ``Bilat: A game-based environment for
  practicing negotiation in a cultural context,'' {\em International Journal of
  Artificial Intelligence in Education}, vol.~19, no.~3, pp.~289--308, 2009.

\bibitem{camerer2003behavioural}
C.~F. Camerer, ``Behavioural studies of strategic thinking in games,'' {\em
  Trends in cognitive sciences}, vol.~7, no.~5, pp.~225--231, 2003.

\bibitem{velmans1991human}
M.~Velmans, ``Is human information processing conscious?,'' {\em Behavioural
  and Brain Sciences}, vol.~14, no.~4, pp.~651--726, 1991.

\bibitem{kihlstrom1996perception}
J.~F. Kihlstrom, ``Perception without awareness of what is perceived, learning
  without awareness of what is learned,'' {\em The science of consciousness},
  pp.~23--46, 1996.

\bibitem{van2012unconscious}
S.~Van~Gaal and V.~A. Lamme, ``Unconscious high-level information processing:
  implication for neurobiological theories of consciousness,'' {\em The
  neuroscientist}, vol.~18, no.~3, pp.~287--301, 2012.

\bibitem{reggia}
J.~A. Reggia, ``The rise of machine consciousness: studying consciousness with
  computational models,'' {\em Neural Netw. 44, 112–131. doi:
  10.1016/j.neunet.2013.03.011}, 2013.

\bibitem{tononi2016integrated}
G.~Tononi, M.~Boly, M.~Massimini, and C.~Koch, ``Integrated information theory:
  from consciousness to its physical substrate,'' {\em Nature Reviews
  Neuroscience}, vol.~17, no.~7, pp.~450--461, 2016.

\bibitem{baars}
B.~Baars, {\em A Cognitive Theory of Consciousness}.
\newblock Cambridge: Cambridge University Press, 1988.

\bibitem{dehaene2017consciousness}
S.~Dehaene, H.~Lau, and S.~Kouider, ``What is consciousness, and could machines
  have it?,'' {\em Science}, vol.~358, no.~6362, pp.~486--492, 2017.

\bibitem{seth2012interoceptive}
A.~K. Seth, K.~Suzuki, and H.~D. Critchley, ``An interoceptive predictive
  coding model of conscious presence,'' {\em Frontiers in psychology}, vol.~2,
  p.~395, 2012.

\bibitem{aleksander}
I.~Aleksander, ``The world in my mind my mind in the world: Key mechanisms of
  consciousness in humans animals and machines.,'' {\em Exeter, Imprint
  Academic}, 2005.

\bibitem{damasio1999feeling}
A.~R. Damasio, {\em The feeling of what happens: Body and emotion in the making
  of consciousness}.
\newblock Houghton Mifflin Harcourt, 1999.

\bibitem{blanke2012multisensory}
O.~Blanke, ``Multisensory brain mechanisms of bodily self-consciousness,'' {\em
  Nature Reviews Neuroscience}, vol.~13, no.~8, pp.~556--571, 2012.

\bibitem{chella}
A.~Chella and R.~Manzotti, ``Machine consciousness: a manifesto for
  robotics.,'' {\em Int. J. Mach. Conscious. 1, 33–51.}, 2009.

\bibitem{manzotti}
R.~Manzotti and A.~Chella, ``Good old-fashioned artificial consciousness and
  the intermediate level fallacy.,'' {\em Frontiers in Robotics and AI, 5,
  39.}, 2018.

\bibitem{james1890principles}
W.~James, F.~Burkhardt, F.~Bowers, and I.~K. Skrupskelis, {\em The principles
  of psychology}, vol.~1.
\newblock Macmillan London, 1890.

\bibitem{nagel1974like}
T.~Nagel, ``What is it like to be a bat?,'' {\em The philosophical review},
  vol.~83, no.~4, pp.~435--450, 1974.

\bibitem{lehar2003world}
S.~M. Lehar, {\em The world in your head: A gestalt view of the mechanism of
  conscious experience}.
\newblock Psychology Press, 2003.

\bibitem{merker1}
B.~Merker, ``From probabilities to percepts a subcortical “global best
  estimate buffer” as locus of phenomenal experience,'' {\em Being in Time:
  Dynamical models of phenomenal experience}, vol.~88, p.~37, 2012.

\bibitem{rudrauf4}
D.~Rudrauf, D.~Bennequin, I.~Granic, G.~Landini, K.~Friston, and K.~Williford,
  ``A mathematical model of embodied consciousness,'' {\em Journal of
  theoretical biology}, vol.~428, pp.~106--131, 2017.

\bibitem{riva1}
G.~Riva, ``The neuroscience of body memory: From the self through the space to
  the others.,'' {\em Cortex, 104, 241-260.}, 2018.

\bibitem{mchugh1}
L.~McHugh and I.~Stewart, {\em The self and perspective taking: Contributions
  and applications from modern behavioral science}.
\newblock New Harbinger Publications, California, 2012.

\bibitem{ciompi1991affects}
L.~Ciompi, ``Affects as central organising and integrating factors a new
  psychosocial/biological model of the psyche,'' {\em The British Journal of
  Psychiatry}, vol.~159, no.~1, pp.~97--105, 1991.

\bibitem{baroncohen3}
S.~Baron-Cohen, ``Joint-attention deficits in autism: Towards a cognitive
  analysis,'' {\em Development and psychopathology}, vol.~1, no.~3,
  pp.~185--189, 1989.

\bibitem{Premack1}
D.~Premack and G.~Woodruff, ``Does the chimpanzee have a theory of mind?,''
  {\em Behavioral and brain sciences}, vol.~1, no.~4, pp.~515--526, 1978.

\bibitem{seth}
A.~K. Seth, ``Consciousness: The last 50 years (and the next),'' {\em Brain and
  neuroscience advances, 2, 2398212818816019.}, 2018.

\bibitem{seth2}
A.~K. Seth and J.~Hohwy, ``Elusive phenomenology, counterfactual awareness, and
  presence without mastery,'' {\em Cognitive neuroscience, 5(2), 127-128.},
  2014.

\bibitem{seth3}
A.~K. Seth, ``Presence, objecthood, and the phenomenology of predictive
  perception,'' {\em Cognitive neuroscience, 6(2-3), 111-117.}, 2015.

\bibitem{revonsuo}
A.~Revonsuo, {\em Consciousness as a Biological Phenomenon}.
\newblock MA: MIT Press, 2005.

\bibitem{crick1990towards}
F.~Crick and C.~Koch, ``Towards a neurobiological theory of consciousness,'' in
  {\em Seminars in the Neurosciences}, vol.~2, p.~203, 1990.

\bibitem{merker2021integrated}
B.~Merker, K.~Williford, and D.~Rudrauf, ``The integrated information theory of
  consciousness: A case of mistaken identity,'' {\em Behavioral and Brain
  Sciences}, pp.~1--72, 2021.

\bibitem{williford}
K.~Williford, D.~Bennequin, K.~Friston, and D.~Rudrauf, ``The projective
  consciousness model and phenomenal selfhood,'' {\em Frontiers in Psychology},
  vol.~9, p.~2571, 2018.

\bibitem{rudrauf5}
D.~Rudrauf, D.~Bennequin, and K.~Williford, ``The moon illusion explained by
  the projective consciousness model,'' {\em Journal of Theoretical Biology},
  vol.~507, p.~110455, 2020.

\bibitem{rudrauf3}
D.~Rudrauf and M.~Debban{\'e}, ``Building a cybernetic model of
  psychopathology: beyond the metaphor,'' {\em Psychological Inquiry}, vol.~29,
  no.~3, pp.~156--164, 2018.

\bibitem{friston3}
K.~Friston, T.~FitzGerald, F.~Rigoli, P.~Schwartenbeck, and G.~Pezzulo,
  ``Active inference: a process theory,'' {\em Neural computation}, vol.~29,
  no.~1, pp.~1--49, 2017.

\bibitem{apps}
M.~A. Apps and M.~Tsakiris, ``The free-energy self: a predictive coding account
  of self-recognition,'' {\em Neuroscience \& Biobehavioral Reviews}, vol.~41,
  pp.~85--97, 2014.

\bibitem{friston2}
K.~Friston, ``The free-energy principle: a unified brain theory?,'' {\em Nature
  reviews neuroscience}, vol.~11, no.~2, pp.~127--138, 2010.

\bibitem{limanowski}
J.~Limanowski and F.~Blankenburg, ``Minimal self-models and the free energy
  principle,'' {\em Frontiers in human neuroscience}, vol.~7, p.~547, 2013.

\bibitem{seth4}
A.~K. Seth, ``The cybernetic bayesian brain.,'' {\em Open MIND. Frankfurt am
  Main: MIND Group.}, 2014.

\bibitem{friston5}
K.~Friston and C.~Frith, ``A duet for one,'' {\em Conscious Cogn 36: 390-405.},
  2015.

\bibitem{constant}
A.~Constant, M.~J. Ramstead, S.~P. Veissiere, J.~O. Campbell, and K.~J.
  Friston, ``A variational approach to niche construction,'' {\em Journal of
  The Royal Society Interface}, vol.~15, no.~141, p.~20170685, 2018.

\bibitem{constant2}
A.~Constant, M.~J. Ramstead, S.~P. Veissi{\`e}re, and K.~Friston, ``Regimes of
  expectations: An active inference model of social conformity and human
  decision making,'' {\em Frontiers in psychology}, vol.~10, p.~679, 2019.

\bibitem{veissiere2020thinking}
S.~P. Veissi{\`e}re, A.~Constant, M.~J. Ramstead, K.~J. Friston, and L.~J.
  Kirmayer, ``Thinking through other minds: A variational approach to cognition
  and culture,'' {\em Behavioral and Brain Sciences}, vol.~43, 2020.

\bibitem{joffily}
M.~Joffily and G.~Coricelli, ``Emotional valence and the free-energy
  principle,'' {\em PLoS Comput Biol}, vol.~9, no.~6, p.~e1003094, 2013.

\bibitem{rudrauf2020role}
D.~Rudrauf, G.~Sergeant-Perthuis, O.~Belli, Y.~Tisserand, and G.~Serugendo,
  ``The role of consciousness in biological cybernetics: emergent adaptive and
  maladaptive behaviours in artificial agents governed by the projective
  consciousness model,'' {\em arXiv preprint arXiv:2012.12963}, 2020.

\bibitem{kalbe1}
E.~Kalbe, F.~Grabenhorst, M.~Brand, J.~Kessler, R.~Hilker, and H.~J.
  Markowitsch, ``Elevated emotional reactivity in affective but not cognitive
  components of theory of mind: A psychophysiological study,'' {\em Journal of
  Neuropsychology}, vol.~1, no.~1, pp.~27--38, 2007.

\bibitem{baroncohen1}
S.~Baron-Cohen, ``Precursors to a theory of mind: Understanding attention in
  others,'' in {\em Natural theories of mind: Evolution, development and
  simulation of everyday mindreading} (A.~Whiten and R.~Byrne, eds.), vol.~1,
  pp.~233--251, Basil Blackwell Oxford, 1991.

\bibitem{wimmer1}
H.~Wimmer and J.~Perner, ``Beliefs about beliefs: Representation and
  constraining function of wrong beliefs in young children's understanding of
  deception,'' {\em Cognition}, vol.~13, no.~1, pp.~103--128, 1983.

\bibitem{lamm}
C.~Lamm, C.~Lamm, C.~D. Batson, and J.~Decety, ``The neural substrate of human
  empathy: effects of perspective-taking and cognitive appraisal,'' {\em
  Journal of cognitive neuroscience, 19(1), 42-58.}, 2007.

\bibitem{berthoz2010spatial}
A.~Berthoz and B.~Thirioux, ``A spatial and perspective change theory of the
  difference between sympathy and empathy,'' {\em Paragrana}, vol.~19, no.~1,
  pp.~32--61, 2010.

\bibitem{gross}
J.~J. Gross, ``The emerging field of emotion regulation: An integrative
  review,'' {\em Review of general psychology, 2(3), 271-299.}, 1988.

\bibitem{clement}
F.~Clément and D.~Dukes, ``The role of interest in the transmission of social
  values,'' {\em Frontiers in Psychology, 4, 349.}, 2013.

\bibitem{gergely}
G.~Gergely and J.~S. Watson, ``The social biofeedback theory of parental
  affect-mirroring: The development of emotional self-awareness and
  self-control,'' {\em Inf. International Journal of Psycho-Analysis, 77,
  1181-1212.}, 1996.

\bibitem{beckes}
L.~Beckes and J.~A. Coan, ``Social baseline theory: The role of social
  proximity in emotion and economy of action.,'' {\em Social and Personality
  Psychology Compass, 5(12), 976-988.}, 2011.

\bibitem{fonagy}
P.~fonagy Fonagy and E.~Allison, ``The role of mentalizing and epistemic trust
  in the therapeutic relationship.,'' {\em Psychotherapy, 51(3), 372.}, 2014.

\bibitem{kalisch}
R.~Kalisch, M.~B. Müller, and O.~Tüscher, ``A conceptual framework for the
  neurobiological study of resilience,'' {\em Behavioral and Brain Sciences},
  vol.~38, 2015.

\bibitem{dawson1}
G.~Dawson, A.~N. Meltzoff, J.~Osterling, J.~Rinaldi, and E.~Brown, ``Children
  with autism fail to orient to naturally occurring social stimuli,'' {\em
  Journal of autism and developmental disorders}, vol.~28, no.~6, pp.~479--485,
  1998.

\bibitem{Koechlin2007}
E.~Koechlin and A.~Hyafil, ``Anterior prefrontal function and the limits of
  human decision-making,'' {\em Science}, vol.~318, pp.~594--598, 10 2007.

\bibitem{hadwin1}
J.~A. Hadwin, P.~Howlin, and S.~Baron-Cohen, {\em Teaching children with autism
  to mind-read: The workbook}.
\newblock John Wiley \& Sons, 2015.

\bibitem{hamilton1}
A.~F. d.~C. Hamilton, R.~Brindley, and U.~Frith, ``Visual perspective taking
  impairment in children with autistic spectrum disorder,'' {\em Cognition},
  vol.~113, no.~1, pp.~37--44, 2009.

\bibitem{baroncohen2}
S.~Baron-Cohen, A.~M. Leslie, U.~Frith, {\em et~al.}, ``Does the autistic child
  have a “theory of mind”,'' {\em Cognition}, vol.~21, no.~1, pp.~37--46,
  1985.

\bibitem{leslie1}
A.~M. Leslie and U.~Frith, ``Autistic children's understanding of seeing,
  knowing and believing,'' {\em British Journal of Developmental Psychology},
  vol.~6, no.~4, pp.~315--324, 1988.

\bibitem{onishi1}
K.~H. Onishi and R.~Baillargeon, ``Do 15-month-old infants understand false
  beliefs?,'' {\em science}, vol.~308, no.~5719, pp.~255--258, 2005.

\bibitem{georgeff1998belief}
M.~Georgeff, B.~Pell, M.~Pollack, M.~Tambe, and M.~Wooldridge, ``The
  belief-desire-intention model of agency,'' in {\em International workshop on
  agent theories, architectures, and languages}, pp.~1--10, Springer, 1998.

\bibitem{broekens2008formal}
J.~Broekens, D.~Degroot, and W.~A. Kosters, ``Formal models of appraisal:
  Theory, specification, and computational model,'' {\em Cognitive Systems
  Research}, vol.~9, no.~3, pp.~173--197, 2008.

\bibitem{Camerer1997}
C.~F. Camerer, ``Progress in behavioral game theory,'' {\em Journal of Economic
  Perspectives}, vol.~11, pp.~167--188, 1997.

\bibitem{Alskaif2015}
T.~Alskaif, M.~G. Zapata, and B.~Bellalta, ``Game theory for energy efficiency
  in wireless sensor networks: Latest trends,'' {\em Journal of Network and
  Computer Applications}, vol.~54, pp.~33--61, 5 2015.

\bibitem{Freire2019}
I.~T. Freire, X.~D. Arsiwalla, J.-Y. Puigbò, and P.~Verschure, ``Modeling
  theory of mind in multi-agent games using adaptive feedback control.''
  arXiv:1905.13225, 2019.

\bibitem{nagel2018neural}
R.~Nagel, A.~Brovelli, F.~Heinemann, and G.~Coricelli, ``Neural mechanisms
  mediating degrees of strategic uncertainty,'' {\em Social Cognitive and
  Affective Neuroscience}, vol.~13, no.~1, pp.~52--62, 2018.

\bibitem{gmytrasiewicz2000rational}
P.~J. Gmytrasiewicz and E.~H. Durfee, ``Rational coordination in multi-agent
  environments,'' {\em Autonomous Agents and Multi-Agent Systems}, vol.~3,
  no.~4, pp.~319--350, 2000.

\bibitem{gmytrasiewicz1998bayesian}
P.~J. Gmytrasiewicz, S.~Noh, and T.~Kellogg, ``Bayesian update of recursive
  agent models,'' {\em User Modeling and User-Adapted Interaction}, vol.~8,
  no.~1, pp.~49--69, 1998.

\bibitem{baker2009action}
C.~L. Baker, R.~Saxe, and J.~B. Tenenbaum, ``Action understanding as inverse
  planning,'' {\em Cognition}, vol.~113, no.~3, pp.~329--349, 2009.

\bibitem{pynadath2005psychsim}
D.~V. Pynadath and S.~C. Marsella, ``Psychsim: Modeling theory of mind with
  decision-theoretic agents,'' in {\em IJCAI}, vol.~5, pp.~1181--1186, 2005.

\bibitem{Yoshida2008}
W.~Yoshida, R.~J. Dolan, and K.~J. Friston, ``Game theory of mind,'' {\em PLoS
  Computational Biology}, vol.~4, 12 2008.

\bibitem{yoshida2010cooperation}
W.~Yoshida, I.~Dziobek, D.~Kliemann, H.~R. Heekeren, K.~J. Friston, and R.~J.
  Dolan, ``Cooperation and heterogeneity of the autistic mind,'' {\em Journal
  of Neuroscience}, vol.~30, no.~26, pp.~8815--8818, 2010.

\bibitem{bombari2015studying}
D.~Bombari, M.~Schmid~Mast, E.~Canadas, and M.~Bachmann, ``Studying social
  interactions through immersive virtual environment technology: virtues,
  pitfalls, and future challenges,'' {\em Frontiers in psychology}, vol.~6,
  p.~869, 2015.

\bibitem{bruneau2015going}
J.~Bruneau, A.-H. Olivier, and J.~Pettre, ``Going through, going around: A
  study on individual avoidance of groups,'' {\em IEEE transactions on
  visualization and computer graphics}, vol.~21, no.~4, pp.~520--528, 2015.

\bibitem{perry2015virtual}
T.~S. Perry, ``Virtual reality goes social,'' {\em IEEE Spectrum}, vol.~53,
  no.~1, pp.~56--57, 2015.

\bibitem{narang2019inferring}
S.~Narang, A.~Best, and D.~Manocha, ``Inferring user intent using bayesian
  theory of mind in shared avatar-agent virtual environments,'' {\em IEEE
  transactions on visualization and computer graphics}, vol.~25, no.~5,
  pp.~2113--2122, 2019.

\bibitem{nguyen2020deep}
T.~T. Nguyen, N.~D. Nguyen, and S.~Nahavandi, ``Deep reinforcement learning for
  multiagent systems: A review of challenges, solutions, and applications,''
  {\em IEEE transactions on cybernetics}, vol.~50, no.~9, pp.~3826--3839, 2020.

\bibitem{tisserand2020real}
Y.~Tisserand, R.~Aylett, M.~Mortillaro, and D.~Rudrauf, ``Real-time simulation
  of virtual humans' emotional facial expressions, harnessing autonomic
  physiological and musculoskeletal control,'' in {\em Proceedings of the 20th
  ACM International Conference on Intelligent Virtual Agents}, pp.~1--8, 2020.

\bibitem{botea2004near}
A.~Botea, M.~M{\"u}ller, and J.~Schaeffer, ``Near optimal hierarchical
  path-finding.,'' {\em J. Game Dev.}, vol.~1, no.~1, pp.~1--30, 2004.

\bibitem{ekman1997face}
R.~Ekman, {\em What the face reveals: Basic and applied studies of spontaneous
  expression using the Facial Action Coding System (FACS)}.
\newblock Oxford University Press, USA, 1997.

\end{thebibliography}
\end{document}